\documentclass[particles,article,submit,pdftex,moreauthors]{Definitions/mdpi} 

\firstpage{1} 
\pubvolume{1}
\issuenum{1}
\articlenumber{0}
\pubyear{2024}
\copyrightyear{2024}
\datereceived{ } 
\daterevised{ } 
\dateaccepted{ } 
\datepublished{ } 
\hreflink{https://doi.org/} 

\usepackage{comment}
\usepackage{amsmath}
\usepackage{amssymb}
\usepackage{listings}
\usepackage{color}
\usepackage{float}
\usepackage{subcaption}
\usepackage{graphicx} 
\usepackage{siunitx}

\nolinenumbers

\Title{ Hadron Identification Prospects With Granular Calorimeters}

\TitleCitation{Hadron Identification Prospects With Granular Calorimeters}

\Author{Andrea De Vita$^{1,2}$\orcidA{}, Abhishek$^{3}$\orcidB{}, Max Aehle$^{6,10}$\orcidI, Muhammad Awais$^{1,2,10}$\orcidC{},
Alessandro Breccia$^2$\orcidP{},
Riccardo Carroccio$^{1,2}$\orcidG{}, Long Chen$^{6,10}$\orcidT{}, Tommaso Dorigo$^{1,4,5,10}$\orcidD{}, Nicolas R.\ Gauger$^{6,10}$\orcidJ{}, Ralf Keidel$^{6,10}$\orcidQ{}, Jan Kieseler$^{8,10}$\orcidL{}, Enrico Lupi$^{1,2}$\orcidE{}, Federico Nardi$^{1,2,9,10}$\orcidN{}, Xuan Tung Nguyen$^{1,6}$\orcidH{}, Fredrik Sandin$^{4,10}$\orcidF{}, Kylian Schmidt$^8$\orcidM{}, Pietro Vischia$^{5,7,10}$ \orcidK{},  Joseph Willmore$^{1}$\orcidO{}}

\AuthorNames{Andrea de Vita, Abhishek, Muhammad Awais, Alessandro Breccia, Riccardo Carroccio, Long Chen, Tommaso Dorigo, Nicolas R. Gauger, Ralf Keidel, Jan Kieseler, Enrico Lupi, Federico Nardi, Xuan Tung Nguyen, Fredrik Sandin, Kylian Schmidt, Pietro Vischia, and Joseph Willmore }

\AuthorCitation{De Vita, A.; Abhishek; Aehle, M; Awais, M.; Breccia, A.; Carroccio, R.; Chen, L; Dorigo, T.; Gauger, N.R.; Keidel, R.; Kieseler, J.; Lupi, E.; Nardi, F.; Nguyen, X.T.; Sandin, F.; Schmidt, K.; Vischia, P.; Willmore, J.}

\address{
$^{1}$ \quad INFN, sezione di Padova - Via F. Marzolo 8, 35131 Padova, Italy \\
$^{2}$ \quad Universit\`a di Padova, dipartimento di Fisica e Astronomia, Via F. Marzolo 8, 35131 Padova, Italy \\
$^{3}$ \quad National Institute of Science Education and Research, Jatni, 752050, India \\
$^{4}$ \quad Lule{\aa} University of Technology, 971 87 Lule\aa, Sweden \\
$^{5}$ \quad Universal Scientific Education and Research Network, Italy \\
$^{6}$ \quad University of Kaiserslautern-Landau (RPTU), Gottlieb-Daimler-Straße, 67663 Kaiserslautern, Germany \\
$^{7}$ \quad Universidad de Oviedo and ICTEA, Spain \\
$^{8}$ \quad Karlsruhe Institute of Technology, 76131 Karlsruhe, Germany \\
$^{9}$ \quad Laboratoire de Physique Clermont Auvergne, 63170 Aubière, France \\
$^{10}$ \quad MODE Collaboration, 
\url{https://mode-collaboration.github.io}\\
$^*$ \quad Corresponding author: \texttt{andrea.de.vita@cern.ch} \\
}

\abstract{
In this work we consider the problem of determining the identity of hadrons at high energies based on the topology of their energy depositions in dense matter, along with the time of the interactions. 
Using GEANT4 simulations of a homogeneous lead tungstate calorimeter with high transverse and longitudinal segmentation, we investigated the discrimination of protons, positive pions, and positive kaons at 100 GeV.
The analysis focuses on the impact of calorimeter granularity by progressively merging detector cells and extracting features like energy deposition patterns and timing information.
Two machine learning approaches, XGBoost and fully connected deep neural networks, were employed to assess the classification performance across particle pairs.
The results indicate that fine segmentation improves particle discrimination, with higher granularity yielding more detailed characterization of energy showers.
Additionally, the results highlight the importance of shower radius, energy fractions, and timing variables in distinguishing particle types.
The XGBoost model demonstrated computational efficiency and interpretability advantages over deep learning for tabular data structures, while achieving similar classification performance.
This motivates further work required to combine high- and low-level feature analysis, {\it e.g.}, using convolutional and graph-based neural networks, and extending the study to a broader range of particle energies and types.
}


\begin{document}


\section{Introduction}

For thirty or more years until the end of the last century, the purpose of hadron calorimeters instrumenting detectors for particle colliders has been invariably the one of determining with the highest possible precision the collective energy of hadronic jets. Although already in the late 1980s a few studies had shown promising results in the improvement of jet energy measurement through the analysis of the interactions of individual particles within the jet cone and the use of momentum information for charged particles provided by tracker measurements~\cite{zeusef}, a fine segmentation of the calorimeter did not appear sufficiently motivated to be worth the added cost and data volume overhead. 
Then, after the turn of the century, two separate advancements in data analysis dramatically changed that paradigm: the demonstration of boosted jet tagging on one side~\cite{PhysRevD.101.056019, dreyerlund,ATL-PHYS-PUB-2023-021,kasieczka2018}, and the success of particle flow techniques on the other~\cite{Sirunyan_2017,THOMSON200925}.

The contrast could not be starker. The estimate of the total energy deposited by a stream of hadrons does not require a calorimeter to be built with high longitudinal or transverse segmentation: other attributes, such as passive material, total depth in interaction lengths, hermeticity, and detection materials and sensors are the main drivers of performance. Instead, the identification of sub-jet components produced within fat jets by the decay of high-mass boosted particles such as $W$, $Z$, and $H$ bosons and top quarks, as well as the detailed accounting of energy deposited by charged and neutral particles within a jet performed by particle flow algorithms, both require high granularity of detection elements within the calorimeter volume.

If we consider broadly the problem of optimally designing a calorimeter for a future collider application, the two recent motivations of high segmentation mentioned above should be considered with care, as the cost of construction and independent readout of a large number of cells can be very high. However, a third element in this equation may then need to be considered, because a high longitudinal and transverse segmentation, coupled with accurate timing of the harvested signals and with precise tracking of charged particles entering the detector, may enable the identification of the particle species producing the energy deposits in localized portions of the detector, through the use of machine learning techniques.

The discrimination of protons, charged pions, and charged kaons through the topology and timing of their energy depositions in a hadron calorimeter is very difficult, and there is a dearth of studies of this topic in the literature; we only know the CALICE attempt in 2015~\cite{calicepid}. However, the general push toward high granularity of today's and tomorrow's hadron calorimeters requires a careful assessment of the ultimate particle identification performance of these instruments. Kaon tagging, for example, may enable the identification of leading kaons in strange quark jets, opening the way to studies of Higgs channels such as $H \to s \bar{s}$ decays; separation of the three dominant charged hadrons also improves the particle flow performance.

In this work we consider the above problem from a rather abstract standpoint: our goal is to try and assess what may be the ultimate discrimination power of a hadron calorimeter for protons, pions, and kaons, if the detector is built with arbitrarily high segmentation; in addition, we aim to assess how that information gets lost if the cell size is progressively increased.
This way signal features that are particularly relevant to consider in order to balance the discrimination power versus data readout rate and computational demands can also be investigated (see, {\it e.g.}, feature sampling in Ref.~\cite{verhelst2015analogdigital}). A quantitative answer to these questions may be very important in informing the design of instruments for future collider applications.

This article is organized as follows: in Sec.~\ref{sec:Simulation} we describe the Monte Carlo simulations we produced as a basis of our studies. In Sec.~\ref{sec:Features} we describe the construction of useful high-level features extracted from energy and time determinations in calorimeter cells. In Sec.~\ref{sec:Methodology} we describe the metrics used to evaluate the performance and the models we used to assess what discrimination is possible with the use of those topological features. Section~\ref{sec:Results} describes the results we obtain from our study. In Sec.~\ref{sec:RelatedWorks} we discuss related works and their connection with our studies. We conclude in Sec.~\ref{sec:Conclusions}.
\section{Simulation And Data Generation}
\label{sec:Simulation}

In order to study the physical processes occurring inside the calorimeter, \texttt{GEANT4} is used, 
specifically employing \texttt{FTFP\_BERT} as the Reference Physics List~\cite{ALLISON2016186,AllisonGeant,AGOSTINELLI2003250}. The simulated primary particles are $p$, $K^+$, and $\pi^+$, each with energy equal to \SI{100}{\GeV}, and are generated at \SI{3}{\meter} from the center of the calorimeter surface.

\begin{figure}[H]
    \centering
    \begin{minipage}{0.68\textwidth}
        \centering
        \includegraphics[width=\textwidth]{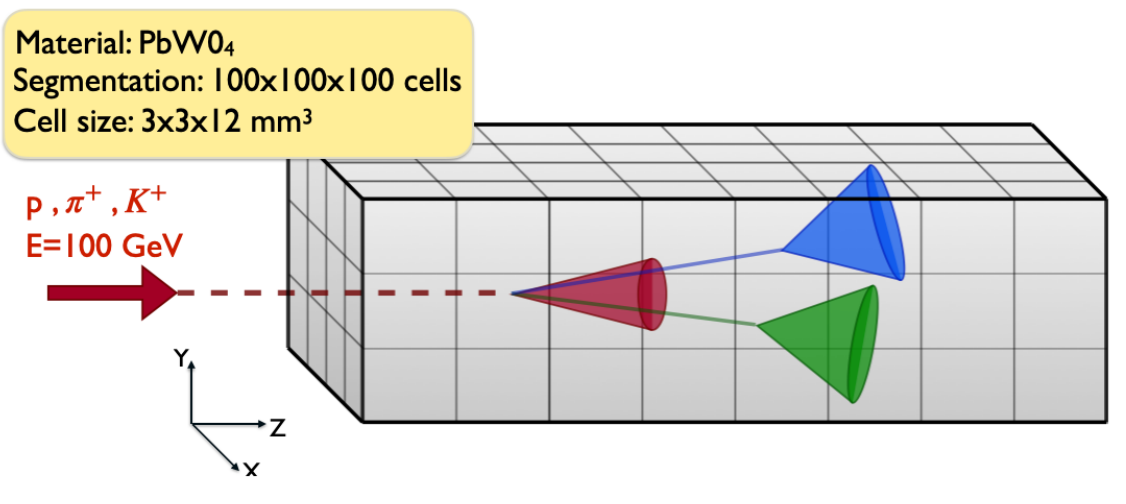}
        \caption*{}
    \end{minipage}
    \\[1em] 
    \begin{minipage}{0.7\textwidth}
        \centering
        \includegraphics[width=\textwidth]{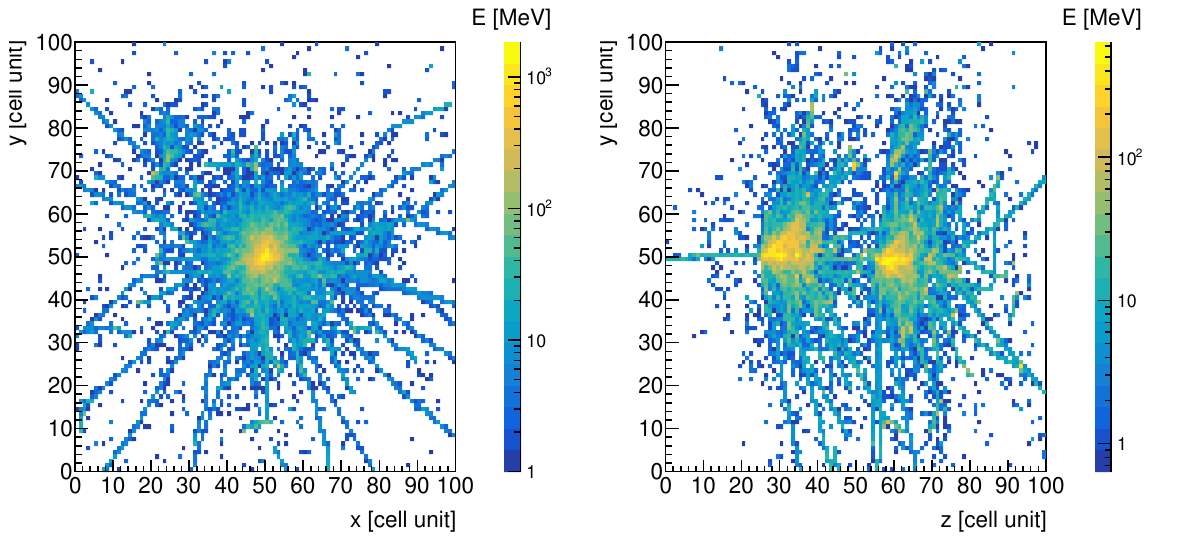}
        \caption*{}
    \end{minipage}
    \caption{(Top) Illustration of the interaction of a simulated charged hadron with a $\text{PbWO}_4$. (Bottom) the energy distribution of a proton shower projected in the XY and ZY planes.}
    \label{fig:simulation}
\end{figure}

The experimental simulated setup consists of a homogeneous calorimeter made up of $100\times100\times100$ cells constructed from Lead Tungstate ($\text{PbWO}_4$) with dimensions of $3 \times 3 \times 12 \, \text{mm}^3$. Therefore, the total size of the calorimeter is $300 \times 300 \times 1200 \, \text{mm}^3$, which corresponds to a lateral width of $7.66 \rho_M$ (Molière radii) and $5.92 \lambda_I$ (interaction lengths), ensuring an average lateral containment of 100\% and a longitudinal containment of approximately 87\% (see Fig.~\ref{fig:simulation}). In each simulated event, the following quantities are extracted through GEANT4's \texttt{SteppingAction}: 
\begin{itemize}
    \item \texttt{PDG index} : This refers to the identity of the particle that released the energy, and its value is encoded according to the Particle Data Group's encoding;
    \item \texttt{PostStep TotalMomentum} : This variable retrieves the total momentum of the particle after it has completed the current step in the simulation;
    \item \texttt{Delta Kinetic Energy} : This variable is computed as the difference between the kinetic energy after and before the GEANT4 simulation step;
    \item \texttt{TotalEnergyDeposit} : This variable retrieves the total energy deposited during the simulation step;
    \item \texttt{PostStep GlobalTime} :  This variable measures the GlobalTime (time since the beginning of the event) after the GEANT4 step. 
    \item  \texttt{Spatial coordinates of the cell that recorded the step} : Each step is recorded by a cell, identified by a pair of indices: the cubelet index (representing a $10\times10\times10$ region in the calorimeter) and the cell index (representing the cell within the cubelet). Both indices range from 0 to 99.
\end{itemize}
Some of the above listed quantities are inaccessible in a real experiment, whereas spatial coordinates, deposited energy, and global time can be considered as a high-accuracy version of the final variables observed in a real experiment.

To limit file size, only steps that satisfy the following conditions are saved:\\
$\texttt{TotalEnergyDeposit} \geq 1 \, \text{keV}$ OR $\texttt{Delta Kinetic Energy} \geq 1 \, \text{keV}$.
For each simulated particle ($p,\pi^+,k^+$), 50,000 events are generated, each with the same initial conditions.  The produced event data are organized into 50 ROOT files, each containing 1000 events, where the information is stored as a \texttt{ColumnWise Ntuple}. This format ensures more efficient space management and speeds up read and write operations.

\subsection{Time Smearing}
\label{sec:smearing}

To mimic a more realistic experimental setup, the recorded \texttt{globalTime} is preprocessed before using it to estimate the temporal variables describing the shower. To account for the finite time resolution of the detector, a smearing time of $\sigma = 30 \text{ ps}$ has been introduced, ensuring that the simulated detector is compatible with current technologies~\cite{TimeCalo}.

\section{Definition Of Sensitive Variables}
\label{sec:Features}

To properly characterise the energy showers in the calorimeter, a set of variables that describe the physical and geometric features of the events can be computed. These variables are either global or local. These global variables represent the general characteristics of the event and they can be seen as the baseline of the system description where no segmentation is used. In contrast, local observables give detailed information (in both the space and the time domain) about the shower due to calorimeter segmentation.

Before describing these variables individual properties can also be extracted from each cell. These properties are then used to compute the descriptive variables.

\subsection{Properties Of Calorimeter Cells}

Each calorimeter cell is characterized by three fundamental properties:
\begin{itemize}
    \item \textbf{Position}: The spatial coordinates of the cell within the calorimeter, which determine its location in the detector geometry.
    \item \textbf{Total absorbed energy}: The total energy deposited in the cell during the event.
    \item \textbf{Cell Characteristic time}: The timing information associated with a cell, defined as the weighted average of the times of all energy depositions within the cell, where the deposited energy serves as the weight:
    \begin{equation}
    t_{\text{cell}} = \frac{\sum_i E^{cell}_i t^{cell}_i}{\sum_i E^{cell}_i} 
    \end{equation}
    Here, $E^{cell}_i$ represents the $i$-th energy deposition within the cell, and $t^{cell}_i$ is the corresponding time. The sum is taken over all the energy depositions within the cell.
\end{itemize}

\subsection{Global variables}

The three global variables considered are the following:
\begin{itemize}
  \item \textbf{Total energy deposited in the calorimeter}.
  \item \textbf{Calorimeter characteristic time}\footnote{This property is computed using the cell properties, and for that reason, it could be considered a local variable. However, its meaning is global, as it corresponds to the mean signal time extracted from a homogeneous calorimeter.}: The timing information associated with the calorimeter, defined as the weighted average of the characteristic times of all cells, where the energy deposited in each cell serves as the weight:
  \begin{equation}
  t_{\text{calo}} = \frac{\sum_{\text{cell}} E_{\text{cell}} t_{\text{cell}}}{\sum_{\text{cell}} E_{\text{cell}}} 
  \end{equation}
  \item \textbf{Time of flight of the particle}: The time of flight (ToF) of the primary particle, hypothetically extracted from a tracker-like detector that is 3 meters long and placed before the calorimeter. Assuming that it is possible to measure the creation time and the arrival time of the particle at the calorimeter interface with perfect resolution, this feature is extracted as follows:
    \begin{align}
        p\;\;\; &= \frac{\sqrt{E^2 - m^2c^4}}{c} \\
        v\;\;\; &= (p/E)*c^2\\
        t_{TOF} &= d/v
    \end{align}
    where $E$ and $m$ are the total energy of the particle and its rest mass, respectively. Here $d$ is the distance traveled by the particle and it is equal to \SI{3}{\meter}; a 30 ps smearing is however added to time measurements later, see {\it infra}, Sec.~\ref{sec:smearing}.
\end{itemize}

\subsection{Local Variables}

The introduction of longitudinal and transverse segmentation in the calorimeter enables the study of the evolution of the energy shower as the particle interacts with the calorimeter. Based on this concept and considering the physical properties of the particles under examination, it is possible to define a set of local variables:
\begin{itemize}
\item \textbf{First nuclear interaction vertex position}: The position of the first nuclear interaction vertex provides an indirect measure of the probability that a particle will interact with the medium through which it is passing. This probability, represented by the particle's nuclear cross section, depends on the properties of the medium, the energy of the particle, and the particle's identity. Therefore, when the first two factors are held constant, the position of the first interaction vertex becomes a variable sensitive to the particle's identity. To determine this position, the First Nuclear Interaction Vertex Finder is used (see Appendix~\ref{sec:appendixA}).

\item \textbf{First interaction vertex time}: The instant at which the first nuclear interaction vertex takes place can be defined as the characteristic time of the cell identified as containing that vertex.

\item \textbf{Speed}: Given the first nuclear interaction vertex position and the first interaction vertex time, the particle speed is defined as the ratio between these two quantities.

\item $\mathbf{\Delta_t}$: Given the first interaction vertex Time $t_V$ and the time when 50\% of the total deposited energy is exceeded ($t_{50}$), it is possible to define $\Delta_t = t_V - t_{50}$.

\item \textbf{Fraction of energy deposited after the first vertex}: Referring to the longitudinal segmentation, the calorimeter can be thought as a collection of layers perpendicular to the direction of the primary particle. Hence, the fraction of energy released after the vertex is defined as the energy released in the calorimeter cells located after the layer containing the vertex.

\item \textbf{Number of non-empty cells before the first vertex layer}: Using the same logic applied to compute the fraction of energy deposited after the first vertex, it is also possible to count the number of non-empty cells in the layers preceding the one containing the vertex.

\item \textbf{Number of non-empty cells}: The total number of cells for which the deposited energy is greater than \SI{0.1}{\MeV}.

\item \textbf{Maximum cell energy}: Maximum cell energy refers to the highest total energy deposited in the cells of the calorimeter.

\item \textbf{Second maximum cell energy}: This variable measures the total energy deposited in the calorimeter cells, representing the second highest value.

\item \textbf{Total energy close to the first vertex and fraction of energy close to the first vertex}: Once the cell of the primary vertex has been identified, it is possible to define a sphere with radius $d$ (cell unit), centered on the selected cell. The total energy deposited in the cells within this sphere represents the total energy close to the first vertex. Thus it can be used to compute the fraction of the total energy deposited in the calorimeter.

\item \textbf{Maximum energy deposited close to the first vertex}: Once the cell of the primary vertex has been identified, it is possible to define a sphere with radius $d$ (cell unit), centered on the selected cell; for different studied segmentations $d$ varies between 2 and 5 cell units. The maximum energy near the primary vertex corresponds to the highest total energy deposited in one of the cells of the sphere.

\item \textbf{Energy variance close to the first vertex}: After identifying the cell which contains the primary vertex, the calorimeter slice that contains that cell can be taken into account. The energy values of individual cells in that layer can then be used to find the variance of the energy deposited in a calorimeter slice that is centered on the cell with the primary vertex.

\item \textbf{Distance between the cell with maximum energy and the first vertex cell}: By considering all the cells of the calorimeter, it is possible to define the distance between the cell containing the primary interaction vertex and the cell with the maximum energy deposition.

\item \textbf{Distance between the maximum energy and second maximum energy cells}: By considering all the cells of the calorimeter, it is possible to define the distance between the cells with the first and second maximum energy depositions.

\item \textbf{Energy close to energy peak and fraction of energy close to energy peak}: After the primary vertex, a peak of deposited energy is generated. The position of this peak can be determined using the energy peak finder (see Appendix~\ref{sec:appendixB}). Similar to the cell containing the primary vertex, a sphere with radius $d$ (cell unit), centered on the cell containing the energy peak, can be defined. This sphere makes it possible to estimate the total energy deposited around the peak and the corresponding fraction of total energy.

\item \textbf{Left and right energy deposition asymmetry}: The impinging position of the primary particle can be considered the center of the reference system, so it is necessary to change the reference system from that of the simulation to the one just described: 
\begin{align}
    x^{\ast} &= x - x_c \\
    y^{\ast} &= y - y_c
\end{align}
Once the new reference system has been adopted, it is possible to compare the left and right energy deposition. There are two methods: the standard definition ($E^{LR}$) and the geometrical definition ($\bar{E}^{LR}$). The former is defined as follows
\begin{align}
    E_x^{LR} &= \sum_{i=1}^N \text{sgn}(x^{\ast}_i) \cdot E_i\quad \text{where } \text{sgn}(x^{\ast}_i) = 
    \begin{cases} 
      1 & \text{if } x^{\ast}_i > 0 \\
      -1 & \text{if } x^{\ast}_i < 0 \\
      0 & \text{if } x^{\ast}_i = 0 
    \end{cases} \\
    E_y^{LR} &= \sum_{i=1}^N \text{sgn}(y^{\ast}_i) \cdot E_i\quad \text{where } \text{sgn}(y^{\ast}_i) = 
    \begin{cases} 
      1 & \text{if } y^{\ast}_i > 0 \\
      -1 & \text{if } y^{\ast}_i < 0 \\
      0 & \text{if } y^{\ast}_i = 0 
    \end{cases} \\
    E^{LR} &= \sqrt{(E_x^{LR})^2 + (E_y^{LR})^2}
\end{align}
The geometrical definition, on the other hand, is defined as the following:
\begin{align}
    \bar{E}_x^{LR} &= \sum_{i=1}^N x^{\ast}_i \cdot E_i \\
    \bar{E}_y^{LR} &= \sum_{i=1}^N y^{\ast}_i \cdot E_i \\
    \bar{E}^{LR} &= \sqrt{(\bar{E}_x^{LR})^2 + (\bar{E}_y^{LR})^2}
\end{align}

\item $\mathbf{R_E^{cell}}$: The energy ratio $R_E^{cell}$ is defined as the following
\begin{equation}
  R_E^{cell} = \frac{E_{max} -  E_{2^{nd} max}}{E_{max} +  E_{2^{nd} max}}
  \end{equation}
  Here, $E_{max}$ represents the maximum total energy deposited in one cell and $E_{2^{nd} max}$ is the second maximum total energy.

\item $\mathbf{\Delta_E^{cell}}$: The energy Delta $\Delta_E^{cell}$ is the numerator of $R_E^{cell}$.

\item $\mathbf{F_E}$: The energy fraction is defined as the following
\begin{equation}
  F_E = \frac{E \text{( within up to $\pm N$ cells around $E_{max}$)}}{E \text{( within up to $1$ cell around $E_{max}$)}} - 1
  \end{equation}
  Here, $N$ can be tuned and it defines a cube around the cell with the maximum total energy.
\end{itemize}

\subsubsection{Physics-Based Observables}
\label{sec:Physics_features}

The slightly lower response of the calorimeter to protons compared to pions of the same energy can be attributed to the fact that, on average, a smaller fraction of the shower energy in proton-induced showers is carried by $\pi^0$-mesons than in pion-induced ones~\cite{AKCHURIN1998380}. This difference arises because of the requirement of baryon-number conservation in nuclear interactions. When a proton undergoes its first nuclear interaction in the absorber material, the most energetic particle produced is typically a baryon~\cite{AKCHURIN1998380}. As this leading particle undergoes subsequent interactions, the most energetic particle produced remains likely to be a baryon. The conservation of the baryon number limits the energy available for the production of $\pi^0$, which generates the calorimeter signal. In contrast, pion-induced showers are not subject to this restriction, which allows more energy to be channeled into $\pi^0$ production~\cite{AKCHURIN1998380}.

The origin of these observed differences between proton and pion showers strongly suggests that the measurable effects are not limited to these particles. In particular, significant differences are also expected between kaon and pion showers. Similarly to baryon number conservation in proton showers, the strangeness quantum number is conserved by strong interactions that occur during kaon-induced showers. The strange (anti-)quark contained in the incident kaon is likely to be transferred to a highly energetic particle during each stage of the shower development~\cite{AKCHURIN1998380}.

The expected outcome is a broader lateral shower profile and a more symmetric signal distribution for protons and kaons compared to pion-induced showers. Furthermore, the electromagnetic fraction is higher for pions than for protons and kaons~\cite{AKCHURIN1998380}.

\begin{itemize}
\item \textbf{Fraction of energy deposited close to the beam axis}: Differences in the fraction of calorimetric signal in the central tower can also be explained by this leading particle effect~\cite{AKCHURIN1998380}. Since the leading particle carries a large fraction of the momentum of the incident particle, it is expected to travel almost in the same direction~\cite{AKCHURIN1998380}. In the case of pion-induced showers, it is most likely that the secondary particle is a $\pi^0$, which means it will generate a large signal in the central calorimeter tower. On the other hand, if the primary particle is a proton, the $\pi^0$’s component of its shower consists of soft $\pi^0$s, which are are produced, on average, at larger angles~\cite{AKCHURIN1998380}. As a result, the lateral profile of the energy deposition by the $\pi^0$s is wider for proton-induced showers than for pion-induced ones. Thus, the fraction of the total signal recorded in the central tower is, on average, smaller for protons and kaons than for pions~\cite{AKCHURIN1998380}.

\item \textbf{Standard spatial observables}: Each energy deposit position can be described by the position of the cell in which it occurred. Thus, it is possible to define the average position along the x, y and z axes of the laboratory reference system ($\bar{x} , \bar{y} , \bar{z}$). With these quantities, the average radius of the energy shower ($R$) and the $\sigma_R$ are the following
\begin{align}
  R &= \frac{\sum_{i=1}^N r_i}{N} \quad \text{with } r_i = \sqrt{(x_i - \bar{x})^2 + (y_i - \bar{y})^2} \\
  \sigma_R &= \sqrt{\frac{\sum_{i=1}^N r_i^2}{N} - R^2} \quad \text{with } r_i = \sqrt{(x_i - \bar{x})^2 + (y_i - \bar{y})^2}
\end{align}
Similarly, the average length of the energy shower ($L$) and its standard deviation are following
\begin{align}
  L &= \frac{\sum_{i=1}^N l_i}{N} \quad \text{with } l_i = z_i - \bar{z} \\
  \sigma_L &= \sqrt{\frac{\sum_{i=1}^N l_i^2}{N} - L^2} \quad \text{with } l_i = z_i - \bar{z}.
\end{align}

\item \textbf{Weighted spatial observables}: Alternatively, the spatial observables can be calculated using the deposited energy as weight. This is how the standard spatial observables are modified once the deposited energy is also taken into account:
\begin{align}
    R^w &= \frac{\sum_{i=1}^N E_i \cdot r_i}{\sum_{i=1}^N E_i} \quad \text{with } r_i = \sqrt{(x_i - \bar{x})^2 + (y_i - \bar{y})^2} \\
    \sigma^w_R &= \sqrt{\frac{\sum_{i=1}^N E_i \cdot r_i^2}{\sum_{i=1}^N E_i} - \left(R^w\right)^2} \quad \text{with } r_i = \sqrt{(x_i - \bar{x})^2 + (y_i - \bar{y})^2} \\
    L^w &= \frac{\sum_{i=1}^N E_i \cdot l_i}{\sum_{i=1}^N E_i} \quad \text{with } l_i = z_i - \bar{z} \\
    \sigma^w_L &= \sqrt{\frac{\sum_{i=1}^N E_i \cdot l_i^2}{\sum_{i=1}^N E_i} - \left(L^w\right)^2} \quad \text{with } l_i = z_i - \bar{z}.
\end{align}

\item $\mathbf{A}$ and $\mathbf{A^w}$:The presence of asymmetries in the transverse profile of the shower can be estimated with the parameters $A$ and $A^w$. Similarly to the left-right energy asymmetry, the impinging position of the primary particle can be considered the center of the reference system. Once the new reference system has been adopted, the parameters $A$ and $A^w$ can be calculated as follows:
\begin{align}
    A &= \sqrt{A_x^2 + A_y^2} \quad \text{with } A_x = \sum_{i=1}^N x_i^{\ast} \quad \text{and} \quad A_y = \sum_{i=1}^N y_i^{\ast} \\
    A^w &= \sqrt{(A^w_x)^2 + (A^w_y)^2} \quad \text{with } A^w_x = \frac{\sum_{i=1}^N (x_i^{\ast} \cdot E_i)}{\sum_{i=1}^N E_i} \quad \text{and} \quad A^w_y = \frac{\sum_{i=1}^N (y_i^{\ast} \cdot E_i)}{\sum_{i=1}^N E_i}
\end{align}

\end{itemize}

\begin{figure}[H]
    \centering
    \setlength{\tabcolsep}{2pt}
    \renewcommand{\arraystretch}{0}
    \begin{tabular}{cc}
        \includegraphics[width=0.5\textwidth]{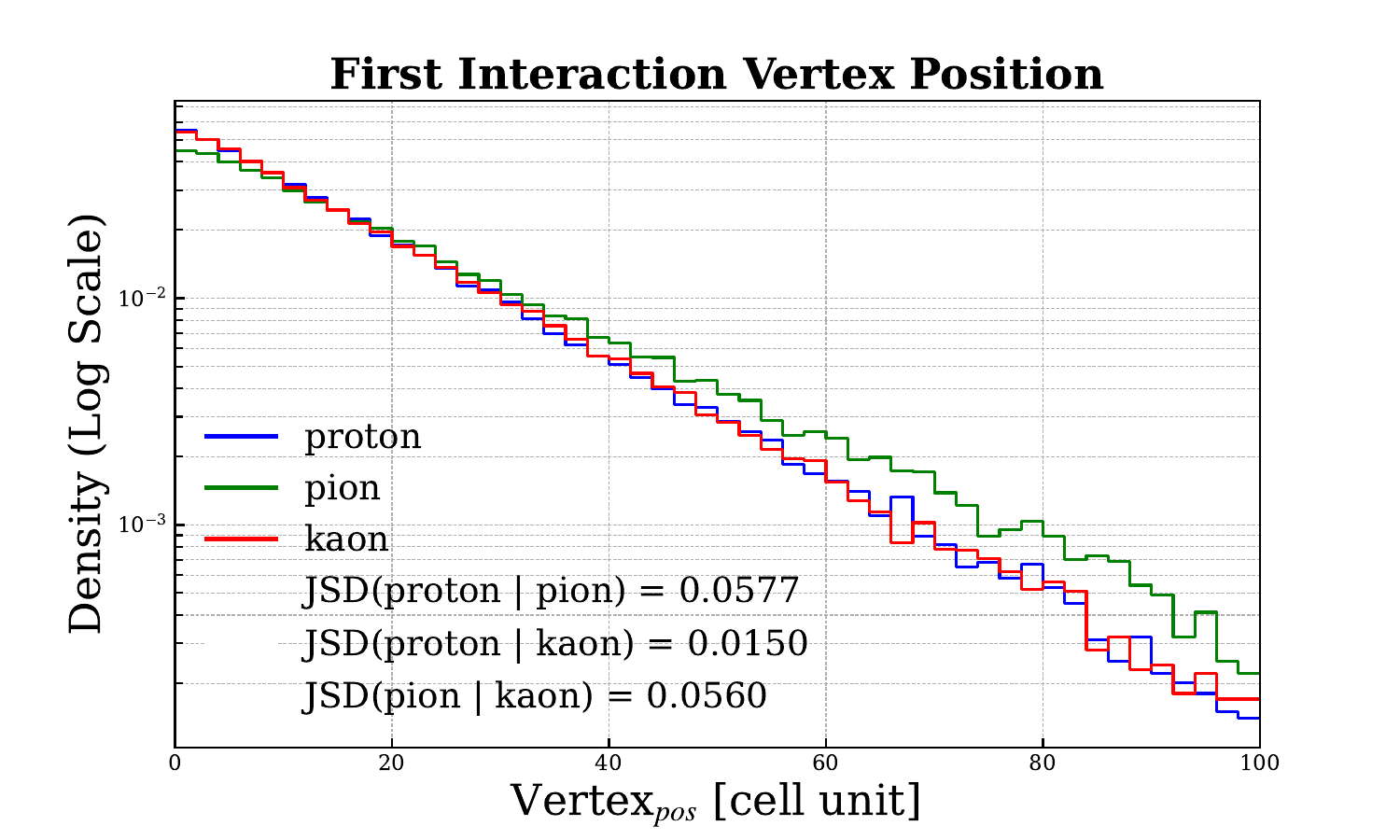} &
        \includegraphics[width=0.5\textwidth]{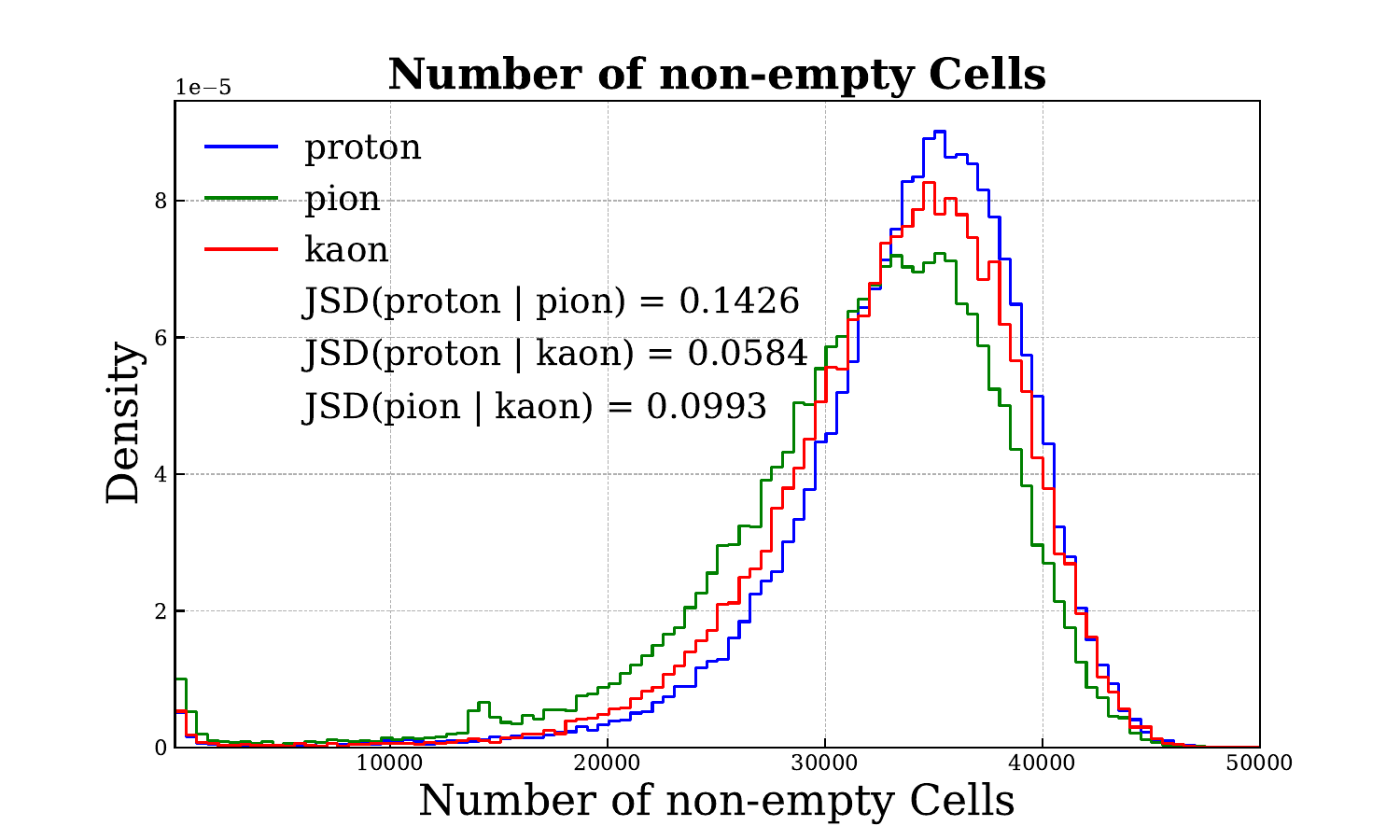} \\
        \includegraphics[width=0.5\textwidth]{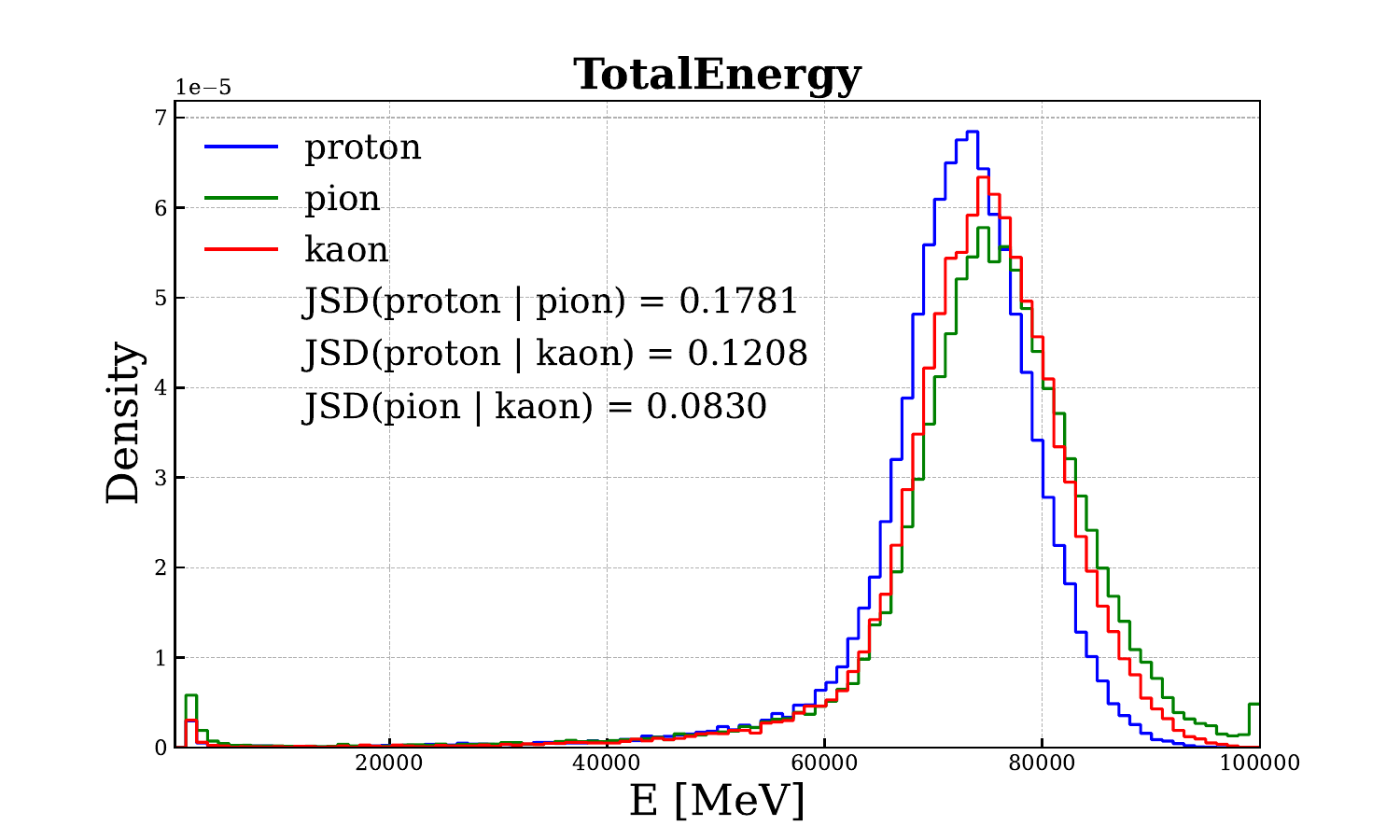} &
        \includegraphics[width=0.5\textwidth]{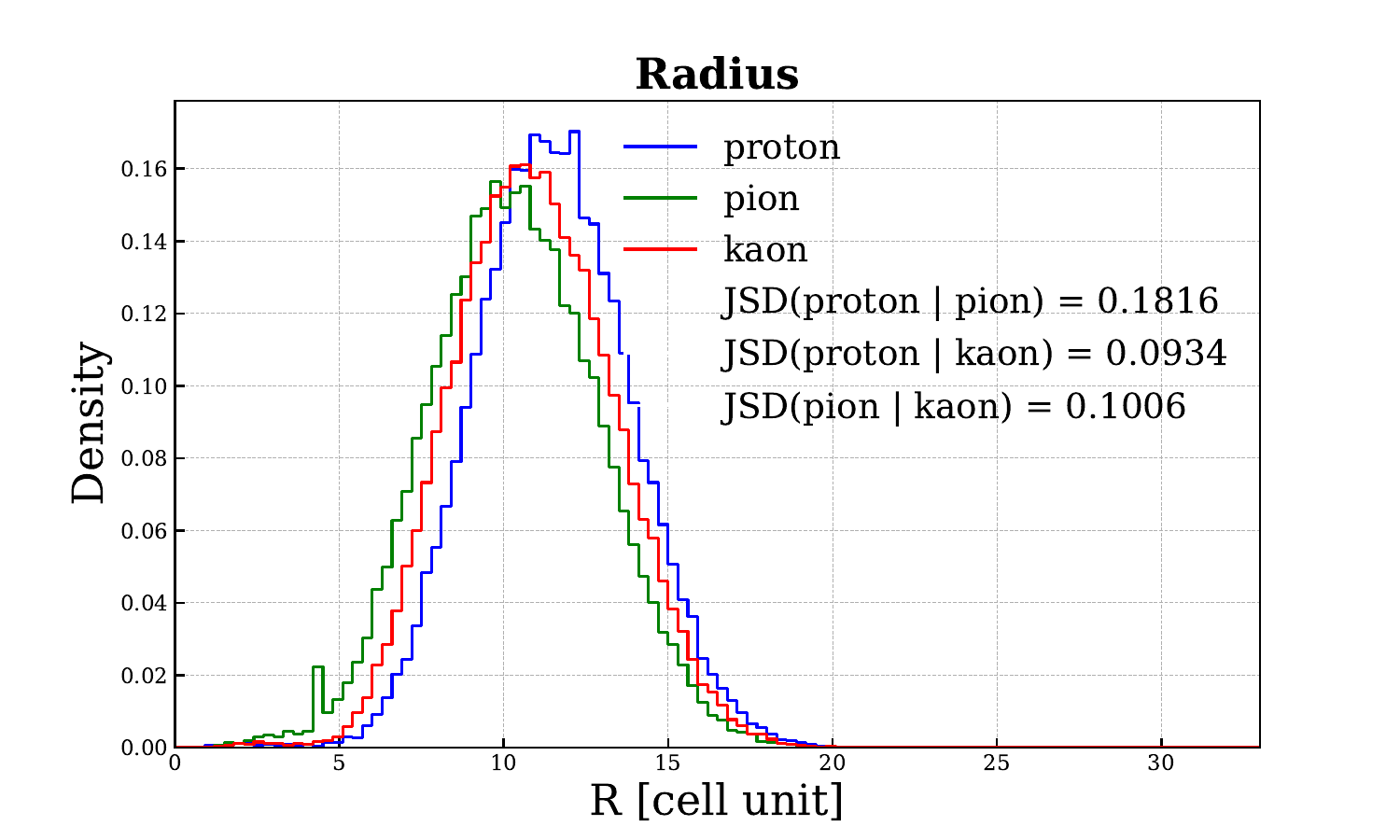} \\
    \end{tabular}
    \caption{Selected feature distributions for proton, pion and kaons (cell size of $3 \times 3 \times 12 \, \text{mm}^3$). For each particle pair is reported also the corresponding Jensen-Shannon divergence to quantify the similarity between their respective distributions. (1) First Nuclear Interaction Vertex Position. (2) Number of non-empty Cells. (3) Total energy deposited in the calorimeter. (4) Radius of the shower.}
    \label{fig:mostImportantFeatures}
\end{figure}
\section{Study Setup And Methodology}
\label{sec:Methodology}

To evaluate the performance of different classifiers, it is crucial to define a set of metrics that assess the model's ability to generalize and make accurate predictions. This section introduces the key metrics used in the study, along with a brief description of each. In addition, the various models tested in this work are presented.

\subsection {Metrics}
\label{sec:metrics}
\begin{itemize}
    \item \textbf{Confusion Matrix:} The confusion matrix is a fundamental tool for evaluating classification models. The confusion matrix provides a foundation for deriving other metrics such as accuracy, precision, recall, and F1-score.

    \item \textbf{ROC Curve:} This is a graphical plot that illustrates the diagnostic ability of a binary classifier as its discrimination threshold is varied. For instance with $p/\pi$ classification, it plots Proton Positive Rate ($PR_p$) against Pion Positive Rate ($PR_{\pi}$), defined as:
    \begin{align*}
        PR_p &= \frac{\text{number of protons classified as protons}}{\text{number of particles classified as protons}}, \\
        PR_{\pi} &= \frac{\text{number of pions classified as pions}}{\text{number of particles classified as pions}}.
    \end{align*}
    The uncertainty associated with the ROC curve is calculated with Wald intervals for the binomial ratio, which is sufficient as the numbers at numerator and denominator are large and the ratio is not close to 0 or 1.

    \item \textbf{Feature Importance:} Feature importance measures the contribution of each input variable to model predictions. For the analysis described in the next section, the gain metric is used in the evaluation of the XGBoost model. This represents the contribution brought to the model by a feature across all splits where the feature is used and it measures how much a particular feature changes the accuracy of the branches it affects.
    
    \item \textbf{Accuracy and Efficiency:} The models which classify showers into particle classes ($p/\pi$, $p/K$, or $\pi/K$) produce pairs of values which sum to 1, and represent the probability an event is in the first or second class. For example, in the classification of protons and pions, the model might output a probability of 0.7 for a proton, meaning the probability for a pion would be 0.3.

    Based on these probabilities, a threshold can be set to assess how far the output of the model can be trusted. Such a threshold reduces the number of the algorithms outputs which are considered reliable, {\it i.e.}, the \textit{efficiency} of the algorithm. But this comes as a trade-off for better \textit{accuracy} (the ability of the model to predict the right class for an input). The accuracy and efficiency curves demonstrate the fluctuation of these metrics with respect to threshold values.

    In addition, the accuracy as a function of the calorimeter cell size is shown. This is done for different configurations, and comparisons are drawn by including uncertainty of values of accuracy. Clopper-Pearson interval are used to estimate this uncertainty, which gives us a confidence interval for a binomial proportion. For an accuracy $a$, computed over $n$ observations with $k$ successes, the confidence interval $[a_\text{low}, a_\text{high}]$ at a given confidence level $1 - \alpha$ is defined as:
    \begin{align}
    a_\text{low} &= \text{BetaInv}\left(\frac{\alpha}{2}, k, n - k + 1\right), \\
    a_\text{high} &= \text{BetaInv}\left(1 - \frac{\alpha}{2}, k + 1, n - k\right),
    \label{eqn:CI}
    \end{align}
    where $\text{BetaInv}(...)$ represents the inverse cumulative distribution function of the Beta distribution.
\end{itemize}

\subsection {Machine Learning Strategy}
\label{sec:ML_architectures}

Two different models were studied to conduct the following study. The first model was built using the XGBoost gradient boosting algorithm, while the second consists of a Fully Connected Deep Neural Network (DNN). Specifically, for each model, a hyperparameter tuning study was conducted using Grid Search.

For each classification task ($p/\pi$, $p/K$, $\pi/K$) the dataset consists of 100k events evenly split between the two particles under examination. It includes the features described in Section~\ref{sec:Features} and it is partitioned into 60\% for training, 20\% for validation, and 20\% for evaluating the model's accuracy.

\subsubsection*{\textbf{XGBoost}}

XGBoost stands for eXtreme Gradient Boosting which is part of the ensemble learning category in the gradient boosting framework. It uses ensemble methods by combining multiple weak learners (decision trees) and building a strong predictor using gradient descent optimization to minimize the errors. It is computationally efficient, manageable to complex relationships, and contains regularization techniques to prevent overfitting.

Because the trees are built one after the other, a technique such as boosting can be used, in which each tree acts to correct errors made by the previous tree, thus allowing it to learn from the updates on those residuals. Boosting uses weak learners (high bias and low predictive power) to create a strong learner that is able to reduce both bias and variance by combining different models. In order to find optimized parameters such as number of trees, learning rate, and tree depth a 3-fold cross-validation has been performed. In addition, the output described in Sec.~\ref{sec:metrics} is obtained by setting the objective function to \texttt{binary:logistic}.

\subsubsection*{\textbf{Deep Neural Network}}

In addition to utilizing XGBoost, a fully connected Deep Neural Network (DNN) is used for classifying protons and pions. However, DNNs often struggle with highly imbalanced datasets and tabular data structures, as highlighted in Grinsztajn {\it et al.}'s findings~\cite{grinsztajn2022treebasedmodelsoutperformdeep}. This limitation arises because DNNs are generally less effective at capturing relationships in tabular data compared to tree-based models.

One effective way to mitigate these issues is through proper data preprocessing, such as standardizing the features. By using Scikit-learn's \texttt{StandardScaler}, one can normalize the dataset to have a mean~($\mu$) of zero and a standard deviation~($\sigma$) of one, ensuring that all features are on a similar scale. This helps DNNs converge more efficiently during training and can significantly improve performance, especially when the dataset is skewed.
\begin{figure}[H]
    \centering
    \includegraphics[width=0.8\textwidth]{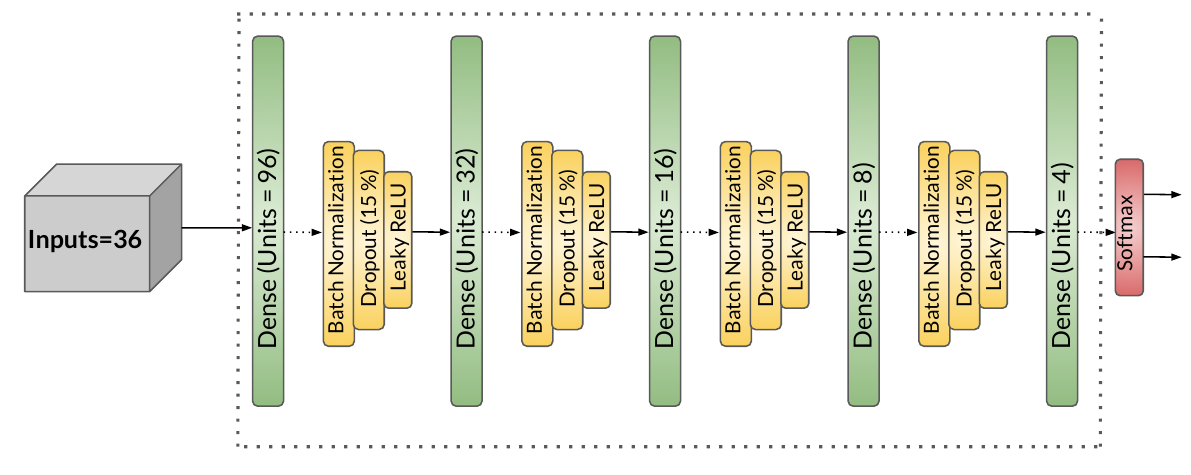} %
    \caption{Deep Neural Network architecture used for classification of hadrons.  }
    \label{fig:NN_arch}
\end{figure}

Figure~\ref{fig:NN_arch} illustrates the architecture utilized for implementing the Neural Network. The network consists of four hidden layers with [96, 32, 16, 4] neurons in each respective layer. Each hidden layer employs a \texttt{LeakyRelu} activation function along with Batch Normalization. Additionally, 15$\%$ random dropout is applied to prevent overfitting. The final  layer output is passed through~\texttt{Softmax} activation function to transform the raw scores into probability scores.\footnote{Note: The \texttt{Softmax} function is not used explicitly here, but is implicitly included in PyTorch's \texttt{CrossEntropyLoss} loss function.}
\begin{table}[H]
\centering
\caption{Main parameters used in the neural network, sourced from PyTorch's library.}
\begin{tabular}{|>{\raggedright\arraybackslash}p{4cm}|c|}
\toprule
\textbf{Parameter} & \textbf{Setting or value} \\ \midrule
Batch size & 128 \\
Type of loss function & \texttt{CrossEntropyLoss} \\
Weight decay & 0.001 \\
Learning rate (initial) & 0.0009 \\
Learning rate (schedule) & $\text{lr} = \frac{\text{lr}}{1 + \text{weight\_decay} \cdot (\text{epoch})^{2.5}}$ \\
Optimizer & \texttt{AdamW} \\
\bottomrule
\end{tabular}

\label{tab:nn_params}
\end{table}

Table~\ref{tab:nn_params} presents the optimal hyperparameters identified through Grid Search. During training, a dynamic learning rate is adopted, which makes learning more stable~\cite{li2019convergencestochasticgradientdescent}.

\section {Results}
\label{sec:Results}

This section presents the results obtained from the considered models. Using the metrics described in the Sec.~\ref{sec:metrics}, performance is evaluated for three distinct classification tasks. In particular, for each simulated particle pair ($p/\pi$, $p/K$, and $\pi/K$), the study investigates the impact of segmentation in a calorimeter versus a homogeneous calorimeter serving as a baseline. In addition, the analysis of the evolution of this contribution as cell size changes is presented.

\subsection{XGBoost}
Using the setup described in Sec.~\ref{sec:ML_architectures}, the results of the study conducted for the three different classification tasks are reported below.

\subsubsection{\texorpdfstring{$p/\pi$ Classification}{p/pi Classification}}
\label{sec:xgboost_ppi}

As explained in Sec.~\ref{sec:Physics_features}, the conservation of baryon number for protons and the dominant branching ratio of neutral pions for charged pions are the key factors used to investigate the differences between showers produced by protons and pions. The consequences are manifested in some of the features, particularly the transverse size of the showers, represented by the radius, and the fraction of energy released along the direction of the interacting particle, {\it i.e.}, the shower core. In addition, the higher radius of protons induces a higher probability of nuclear interaction with the calorimeter material than for pions. It follows that on average a smaller fraction of the pion energy will be deposited in the calorimeter because 100\% containment of charged pions would require a larger longitudinal dimension of the calorimeter.
\begin{figure}[H]
    \centering
    \includegraphics[width=0.53\textwidth]{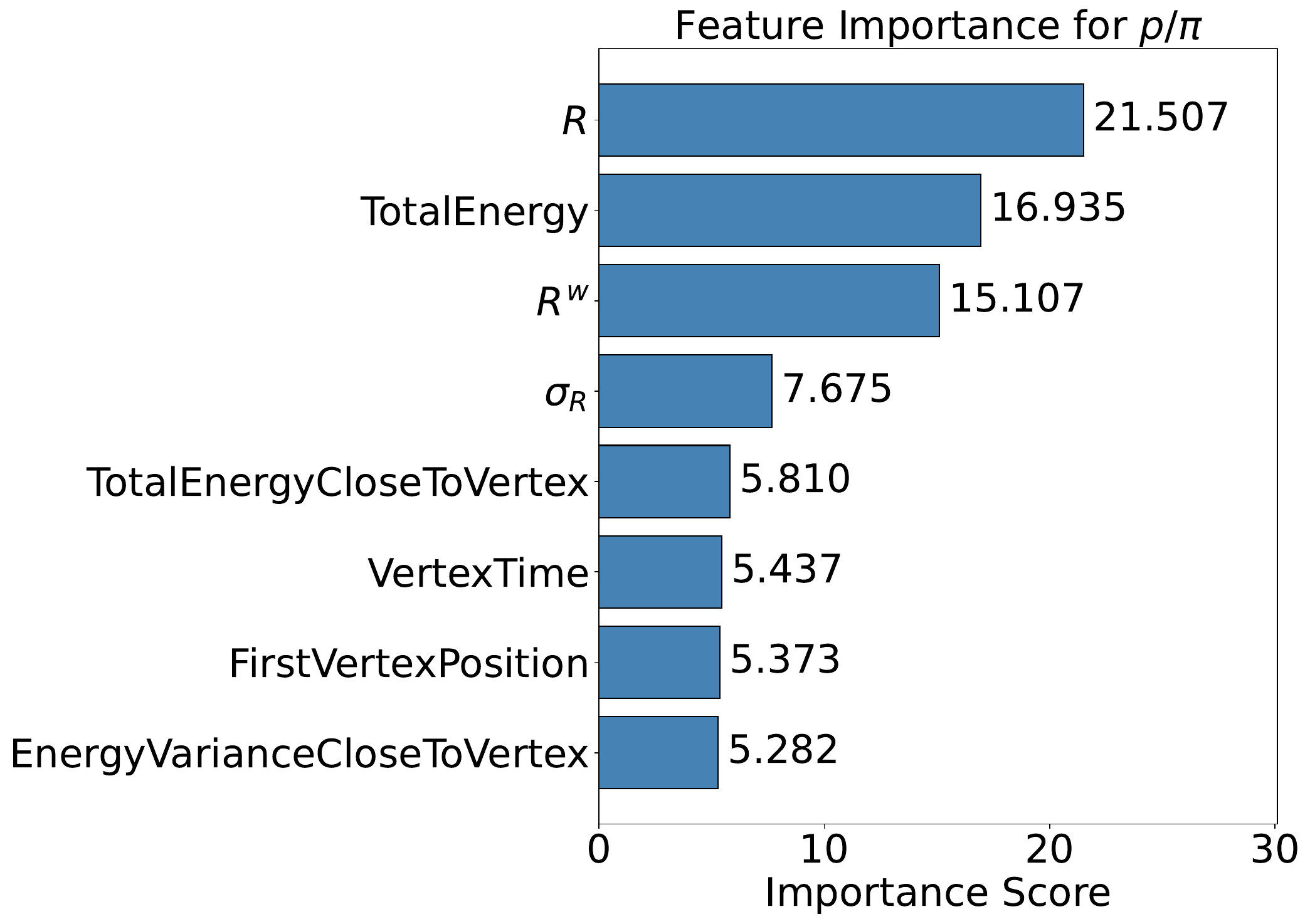} %
    \caption{Top 8 most important features used by the XGBoost model in the $p/\pi$ classification (cell size of $3 \times 3 \times 12 , \text{mm}^3$), ranked in descending order of significance. Each bar represents a feature, with its length proportional to its contribution, measured using the "gain" metric.}
    \label{fig:feature_importance_ppi}
\end{figure}
Table of feature importance in Fig.~\ref{fig:feature_importance_ppi} confirms these assumptions: In addition to total energy deposited inside the calorimeter, the transverse shower profile is another important feature. On the other hand, the vertex position does not seem to contribute seriously. This may be because this quantity is only sensitive in a $\text{PbWO}_4$ calorimeter where longitudinal dimension cells are less than 12 mm. Moreover, the exact position of an interaction vertex is not directly observable and can only be reconstructed through an algorithm which is typically only provided with a lower than 100\% accuracy. So in order for this feature to have greater importance, the cell dimension should be smaller. Also, the accuracy of the First Nuclear Interaction Vertex Finder can be improved. This is an example that demonstrates that the physical insights offered by the XGBoost model are a strong advantage in its use. Such an interpretation helps not only to understand the processes but also the needs which should be considered in future detector designs.

\begin{figure}[t]
    \centering
    \includegraphics[width=.63\textwidth]{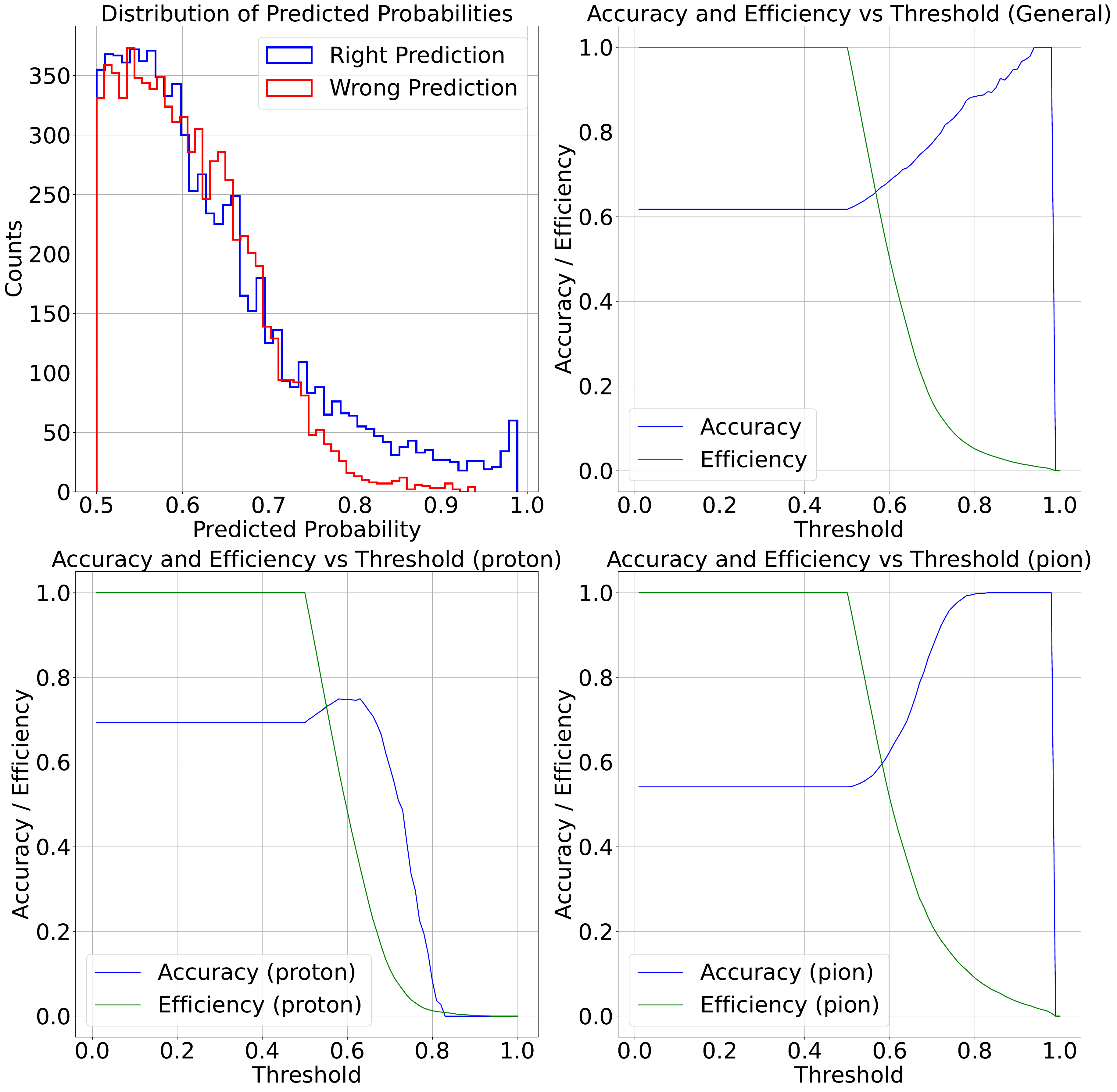} %
    \caption{Analysis of the winning probability and its impact on accuracy and efficiency values (cell size of $3 \times 3 \times 12 , \text{mm}^3$ with XGBoost). (1) Distribution of the winning probability, categorized into winning probabilities with correct classification (in blue) and winning probabilities with incorrect classification (in red). (2) Accuracy and efficiency curves considering both protons and pions. (3) Accuracy and efficiency curves for protons. (4) Accuracy and efficiency curves for pions.}\label{fig:acc_eff_combined}
\end{figure}

Another significant result involves the analysis of the model’s output. As described in Sec.~\ref{sec:metrics}, the output represents two values corresponding to probabilities of the sample being a proton and a pion, respectively. By taking maximum of two probabilities, {\it i.e.} \textit{winning probability}, two distributions can be built. The first one is the winning probability when the sample gets classified correctly and the second is the winning probability when the sample is misclassified. Therefore, these values are on average greater than 0.5 and in Fig.~\ref{fig:acc_eff_combined} it can bee seen that above a certain winning probability threshold, the model returns the correct class most of the time. As described in Sec.~\ref{sec:metrics}, the output can be interpreted as a pair of values representing the probabilities that the sample is a proton or a pion, respectively. Two distributions can be constructed by taking the higher of the two probabilities,{\it i.e.} \textit{winning probability}. One distribution corresponds to the winning probability when the classification is correct, while the other represents the winning probability when the sample is misclassified. In Fig.~\ref{fig:acc_eff_combined}, it can be seen that these values are, by definition, greater than 0.5. Moreover, above a certain winning probability threshold, the model consistently returns the correct class. This suggests that it is possible to define a confidence level that the model must meet to produce reliable output.

\begin{figure}[H]
    \centering
    \begin{subfigure}[t]{0.33\textwidth}
        \centering
        \includegraphics[width=\textwidth]{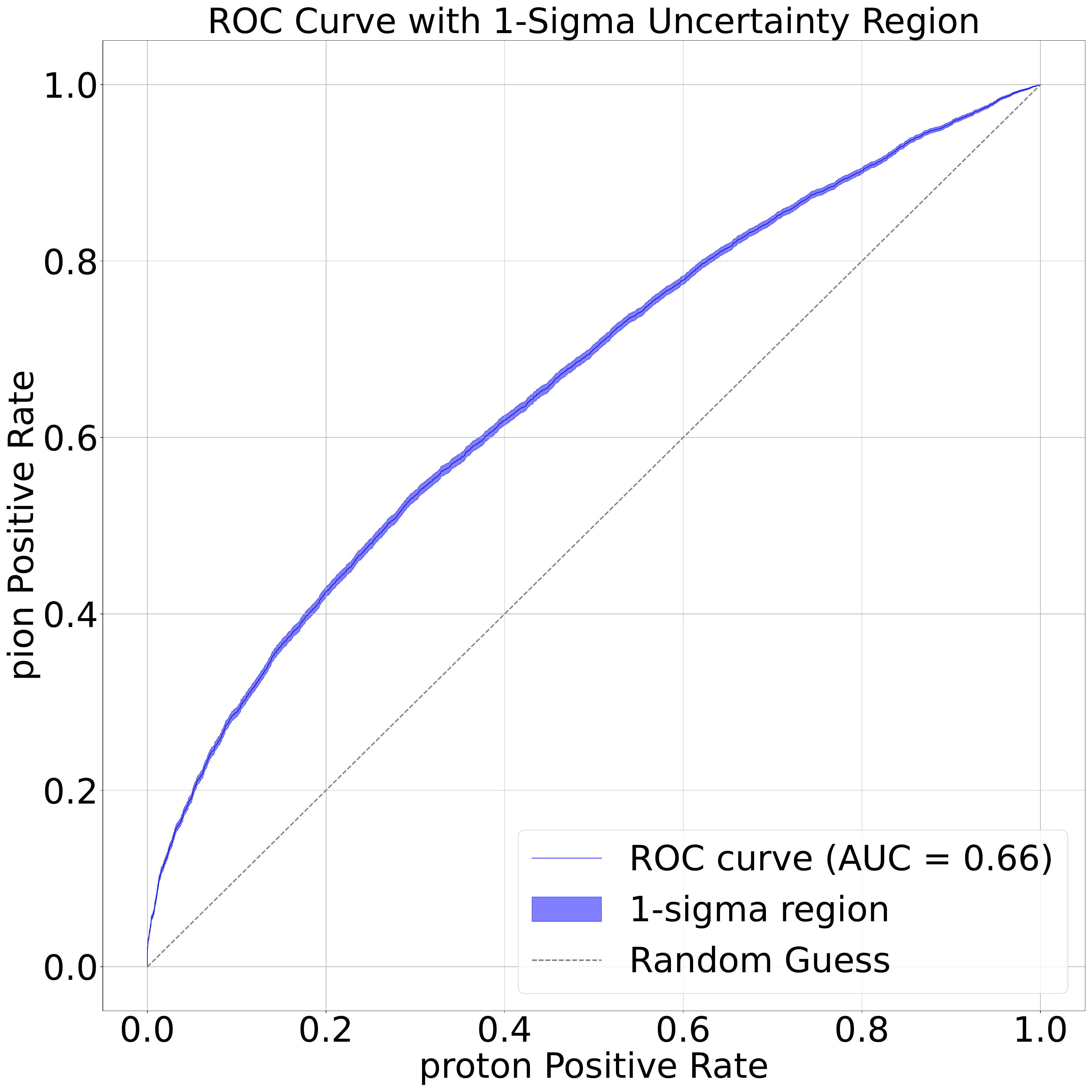}
        \caption*{}
        \label{fig:roc_ppi}
    \end{subfigure}%
    \hfill
    \begin{subfigure}[t]{0.65\textwidth}
        \centering
        \includegraphics[width=\textwidth]{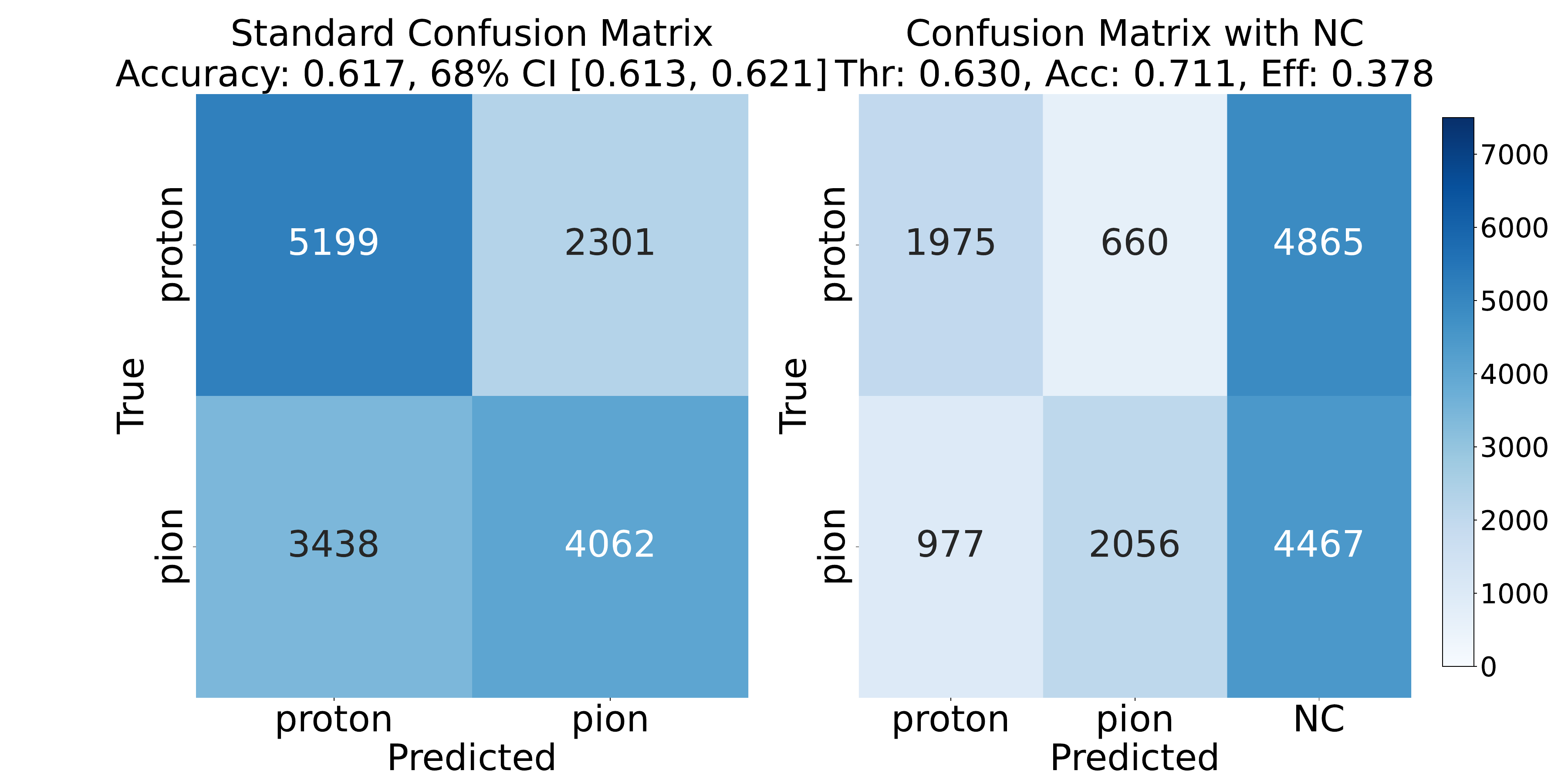}
        \caption*{}
        \label{fig:confusion_matric_ppi}
    \end{subfigure}
    \caption{(Left) $p/\pi$ classification ROC curve. The uncertainty region is computed as described in Sec.~\ref{sec:metrics}. (Right) Confusion matrices illustrating XGBoost's performance with a cell size of $3 \times 3 \times 12 , \text{mm}^3$: confusion matrix at full efficiency and when the threshold on the model's output is set to the value corresponding to the highest achievable accuracy for protons. Particles whose classification confidence falls below the reliability threshold are labeled as 'Not Classified' (NC).}
    \label{fig:confusion_matric_roc_ppi}
\end{figure}

Fig.~\ref{fig:acc_eff_combined} also shows the accuracy and efficiency curves described in Sec.~\ref{sec:metrics}. Increasing the confidence level reduces the model efficiency but simultaneously improves its accuracy. In particular, this effect can be seen in the pion curves, where the accuracy reaches 100\%, meaning that they are more easily discriminated.
The analysis proceeds by evaluating the accuracy achieved in discriminating protons from pions. The right plot of Fig.~\ref{fig:confusion_matric_roc_ppi} presents two confusion matrices. The first matrix illustrates the maximum accuracy achieved at full efficiency, with the corresponding ROC curve shown on the left side of the same figure. The second matrix represents the accuracy obtained when the threshold is set to the value that yields the highest achievable accuracy for protons.

\begin{figure}[H]
    \centering
    \includegraphics[width=0.8\textwidth]{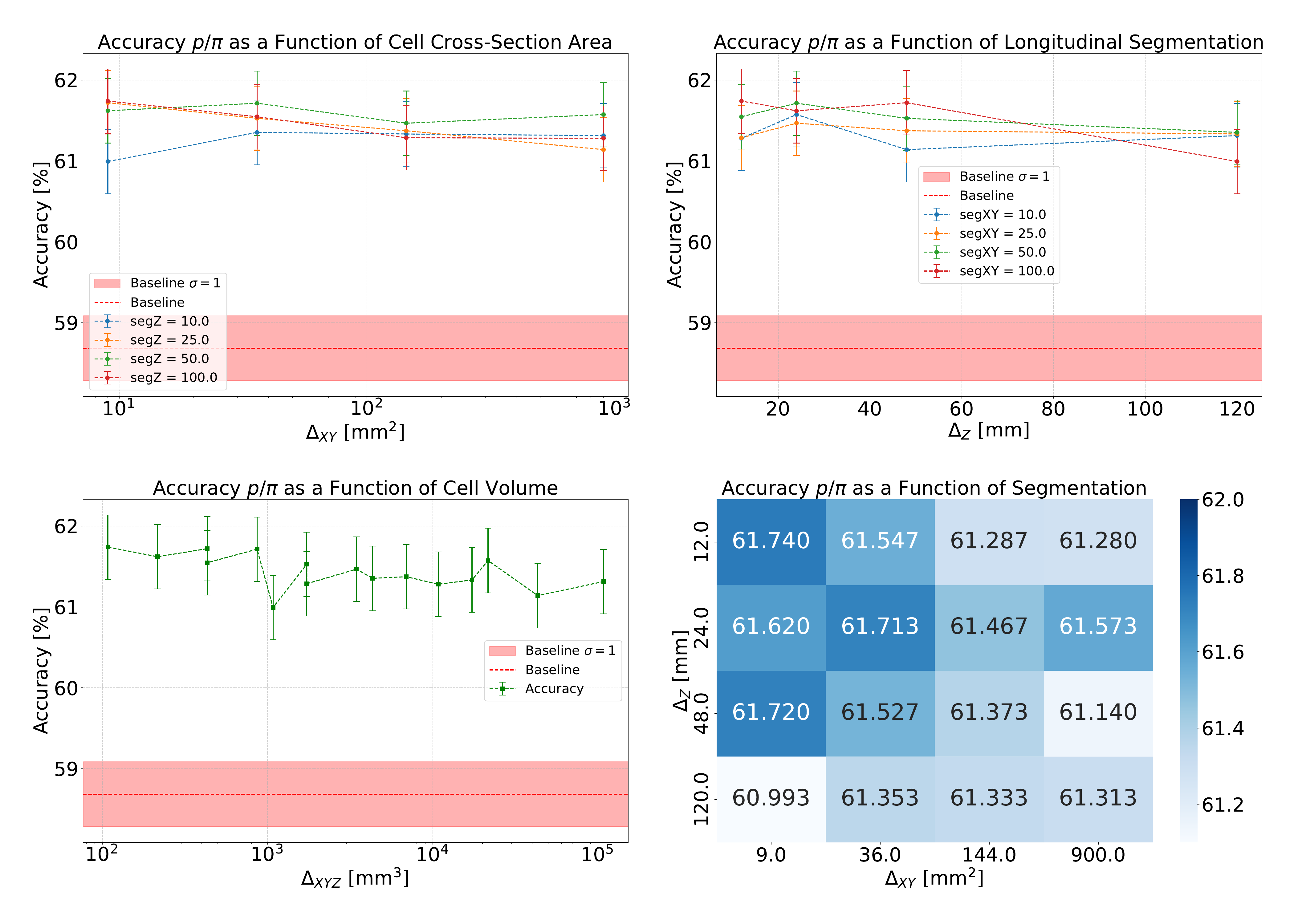} %
    \caption{Summary of results for $p/\pi$ classification with XGBoost. (Top-Left) Dependence of accuracy on cell cross section for different longitudinal segmentations. (Top-Right) Dependence of accuracy on cell length section for different transverse segmentations.  (Bottom-Left) Dependence of accuracy on cell volume. (Bottom-Right) Accuracy for different longitudinal and transverse segmentation configurations. Because of the use of the same input data, the reported accuracy values are correlated. }
    \label{fig:moneyplot_p_pi_xgboost}
\end{figure}

The analysis concludes with a study on the impact of segmentation compared to the use of a homogeneous, non-segmented calorimeter and the effect of cell size on the model's performance. In the top-left panel of Fig.~\ref{fig:moneyplot_p_pi_xgboost}, the accuracy is shown as a function of the cell cross-section for various longitudinal segmentations, while in the top-right panel, the accuracy is plotted against the longitudinal segmentation size for different transverse segmentations. Both plots highlight an improvement in performance with the introduction of segmentation, increasing the baseline accuracy from 58.7\% to an average value of 61.4\%.

In the bottom-left panel of Fig.~\ref{fig:moneyplot_p_pi_xgboost}, the dependence of accuracy on the cell volume shows a decreasing trend as the cell volume increases. This result confirms that introducing smaller volume cells can provide a better description of showers within the calorimeter. In all the plots of accuracy versus cell dimensions, the quoted accuracy values are strongly correlated because of the use of the same input data for training, validation, and testing. Therefore the variability shown by the accuracy results over cell dimensions is more significant than what the uncertainty bars seem to imply. Finally, the bottom right panel of Fig.~\ref{fig:moneyplot_p_pi_xgboost} shows the accuracy obtained for various configurations of longitudinal and transverse segmentations.

\subsubsection{\texorpdfstring{$\pi/K$ Classification}{pi/K Classification}}
\label{sec:xgboost_pik}

Similarly to the $p/ \pi$ classification case discussed above, there are physical reasons that could potentially create a difference between a shower produced by a charged pion and one produced by a charged kaon. Just as the baryon number is conserved in proton showers, the strangeness quantum number is conserved in the strong interactions occurring in kaon-induced showers. The strange (anti-)quark contained in the incident particle is likely to be transferred to a highly energetic particle in each generation of the shower development.  The production of $\pi^0$s in kaon showers is therefore limited by a mechanism very similar to that in proton showers. This can lead to showers that are wider and more symmetric compared to those produced by pions.

\begin{figure}[t]
    \centering
    \begin{subfigure}[t]{0.52\textwidth}
        \centering
        \includegraphics[width=\textwidth]{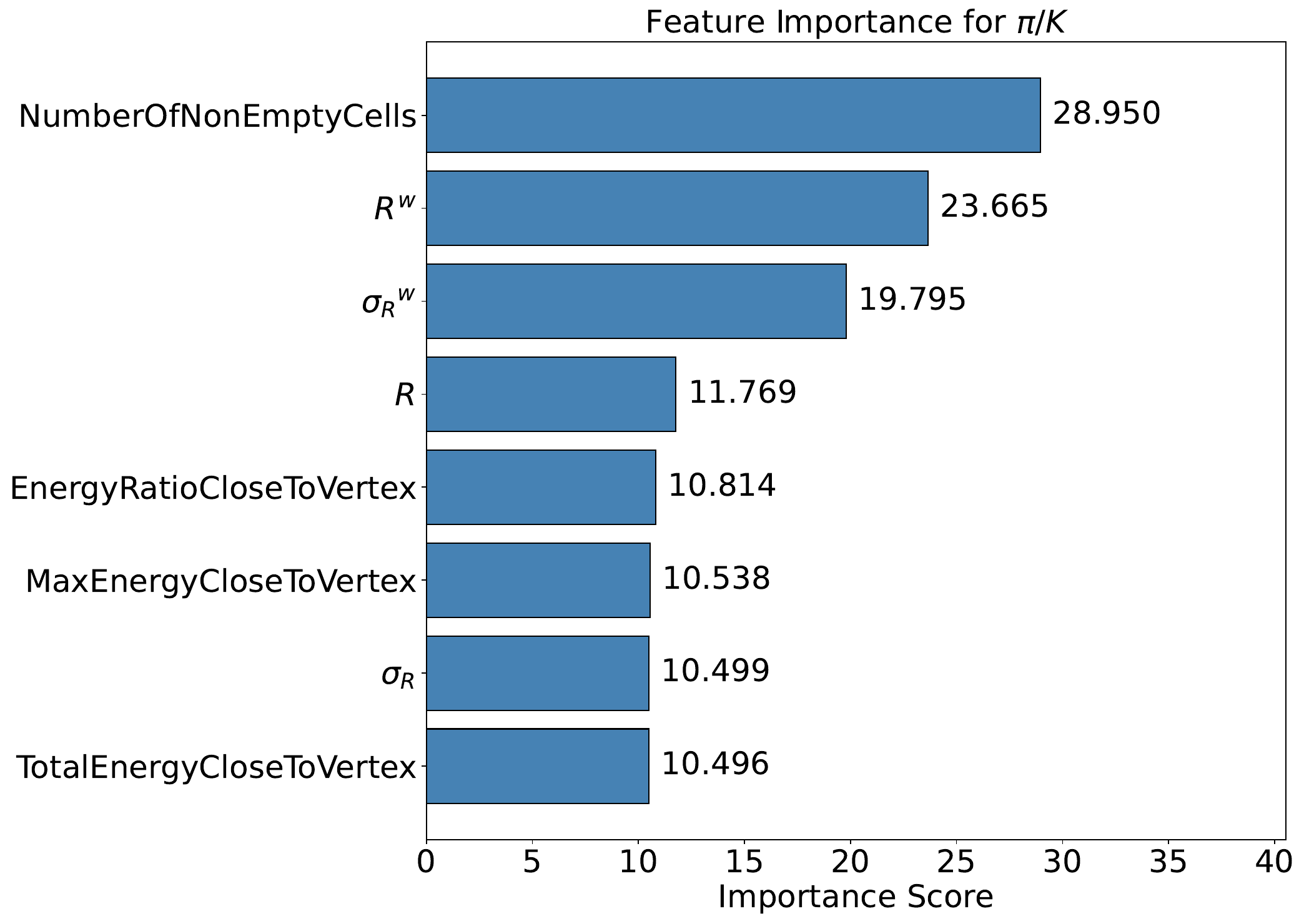}
        \caption*{}
        \label{fig:feature_importance_pik}
    \end{subfigure}%
    \hfill
    \begin{subfigure}[t]{0.41\textwidth}
        \centering
        \includegraphics[width=\textwidth]{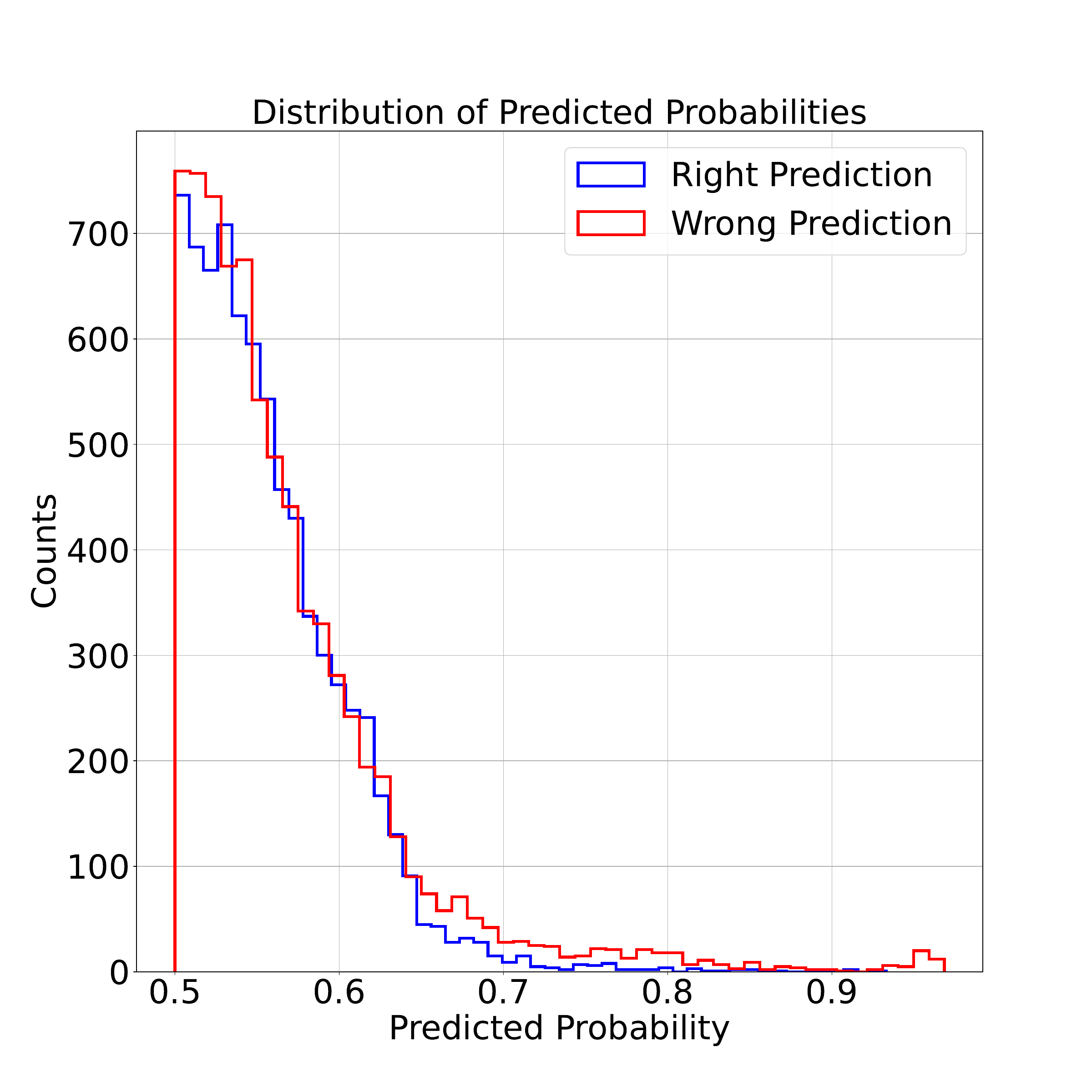}
        \caption*{}
        \label{fig:winning_prob_pik}
    \end{subfigure}
    \caption{Analysis with the XGBoost model for the $\pi/K$ classification (cell size of $3 \times 3 \times 12 \, \text{mm}^3$). (Left) Top 8 most important features used by the XGBoost model in the $\pi/K$ classification, ranked in descending order of feature importance with respect to the gain metric. (Right) Analysis of the winning probability and its impact on accuracy and efficiency values. Distribution of the winning probability, categorized into correct classification (blue) and incorrect classification (red).}
    \label{fig:combined_pik}
\end{figure}

As highlighted in Fig.~\ref{fig:combined_pik} on the left, this is confirmed by the presence of variables describing the transverse development of the shower, such as the radius, among the top-ranking features. It is worth noting the presence of the number of non-empty cells, which indicates a difference between the two types of showers. In particular, looking at Fig.~\ref{fig:mostImportantFeatures} it can be seen that the number of non-empty cells for pions is, on average, smaller than that for kaons.

In the analysis of the winning probability distributions, it becomes clear that the principle outlined in Sec.~\ref{sec:xgboost_ppi} does not apply here, as no threshold exists where correct predictions consistently outnumber incorrect ones (see Fig.~\ref{fig:combined_pik} on the right).

\begin{figure}[H]
    \centering
    \includegraphics[width=0.8\textwidth]{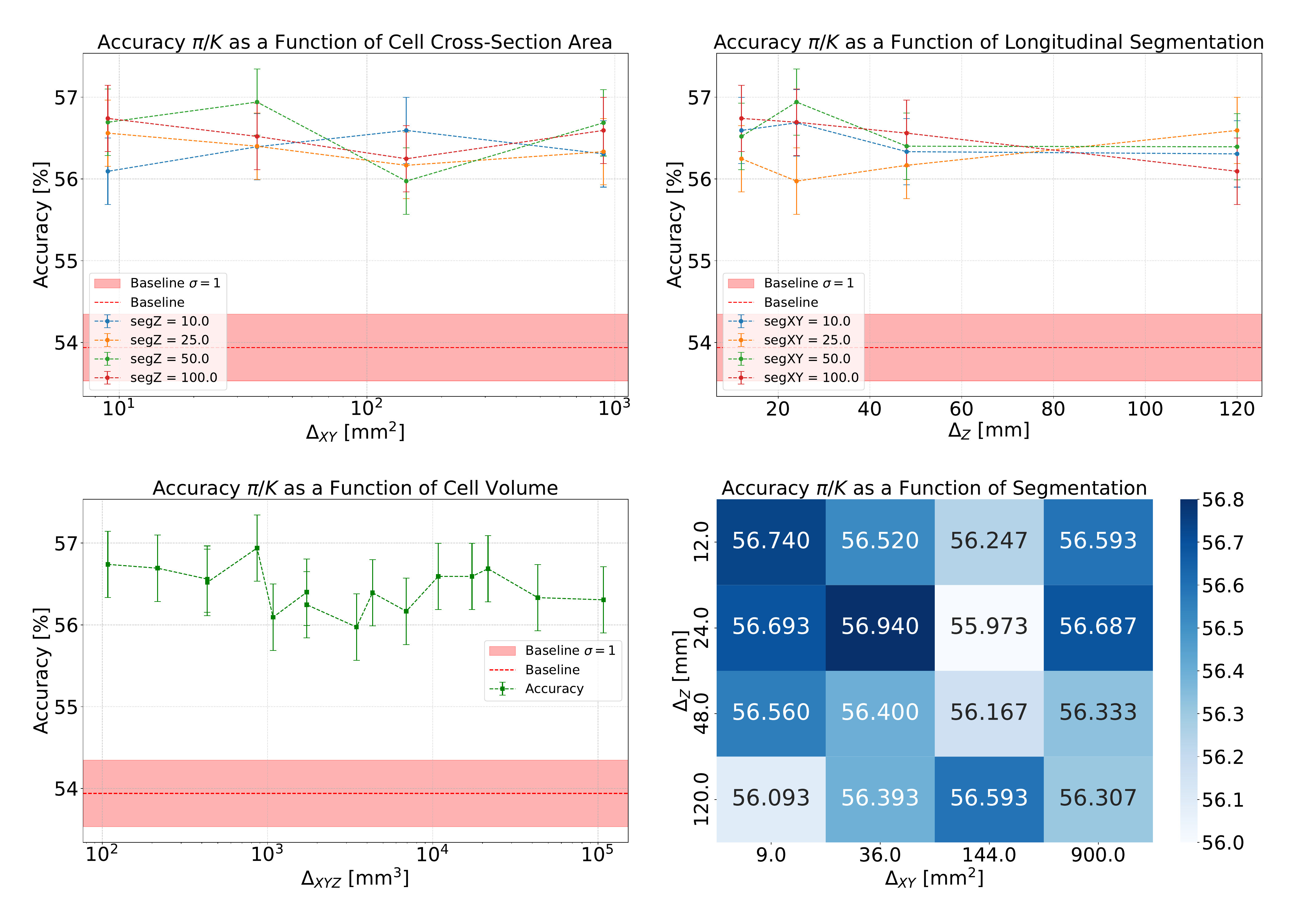}
    \caption{Summary panels for $\pi/K$ classification with XGBoost. (Top-Left) Dependence of accuracy on cell cross section for different longitudinal segmentations. (Top-Right) Dependence of accuracy on cell length section for different transverse segmentations.  (Bottom-Left) Dependence of accuracy on cell volume. (Bottom-Right) Accuracy for different longitudinal and transverse segmentation configurations.  Because of the use of the same input data, the reported accuracy values are correlated. }
    \label{fig:moneyplot_pik_xgboost}
\end{figure}

Finally, observing the contribution of segmentation to the particle identification power \(\pi / K\), an improvement is noted in this case as well (see Fig.~\ref{fig:moneyplot_pik_xgboost}, top). On the other hand, the trend with respect to cell volume does not seem to follow a defined pattern, suggesting that exploring smaller cell sizes might be necessary to fully resolve the showers induced by pions and kaons (see Fig.~\ref{fig:moneyplot_pik_xgboost}, bottom left).

\subsubsection{\texorpdfstring{$p/K$ Classification}{p/K Classification}}
\label{sec:xgboost_pk}

\begin{figure}[H]
    \centering
    \begin{subfigure}[b]{0.43\textwidth}
        \centering
        \includegraphics[width=\textwidth]{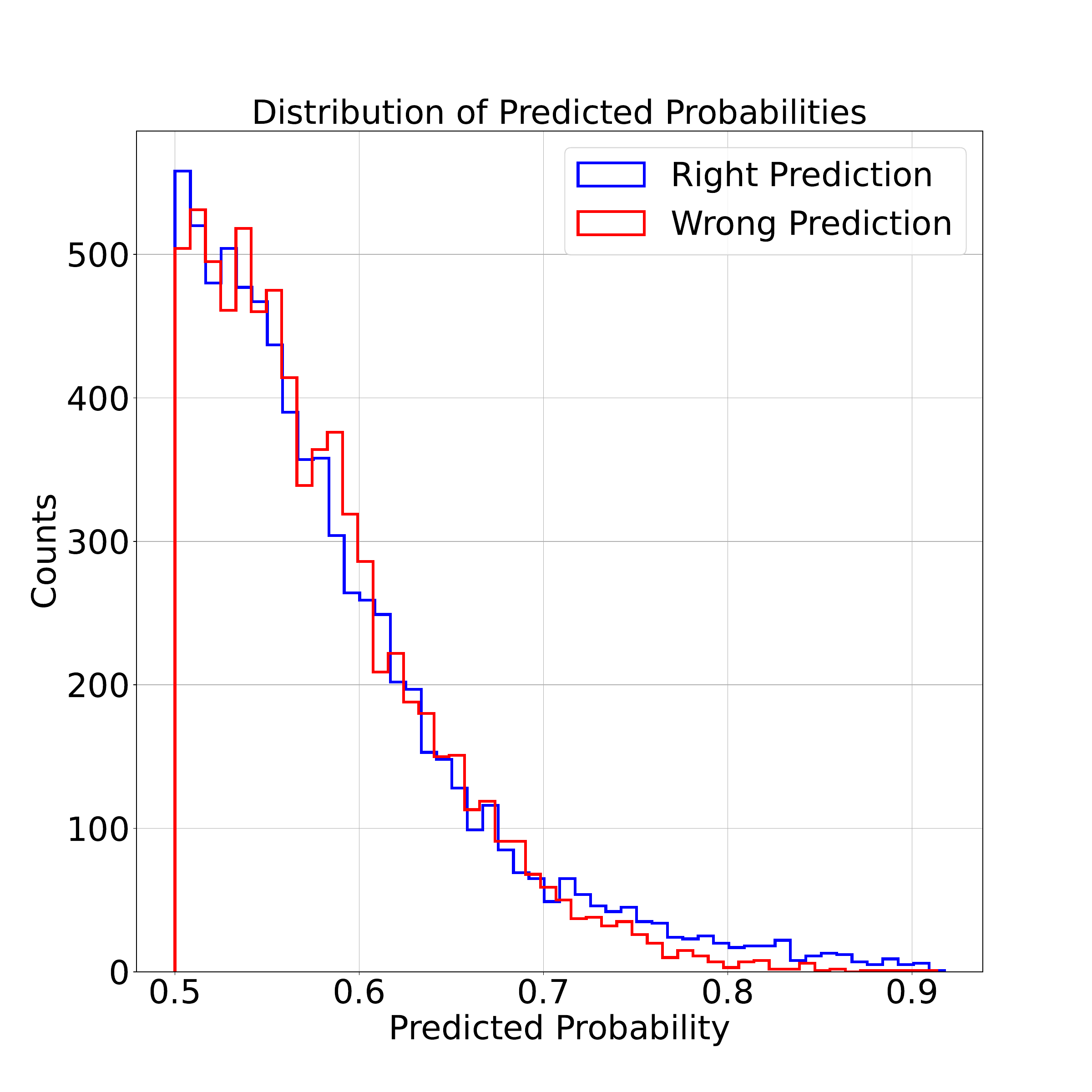}
        \caption*{}
        \label{fig:winning_prob_pk}
    \end{subfigure}
    \hfill
    \begin{subfigure}[b]{0.545\textwidth}
        \centering
        \includegraphics[width=\textwidth]{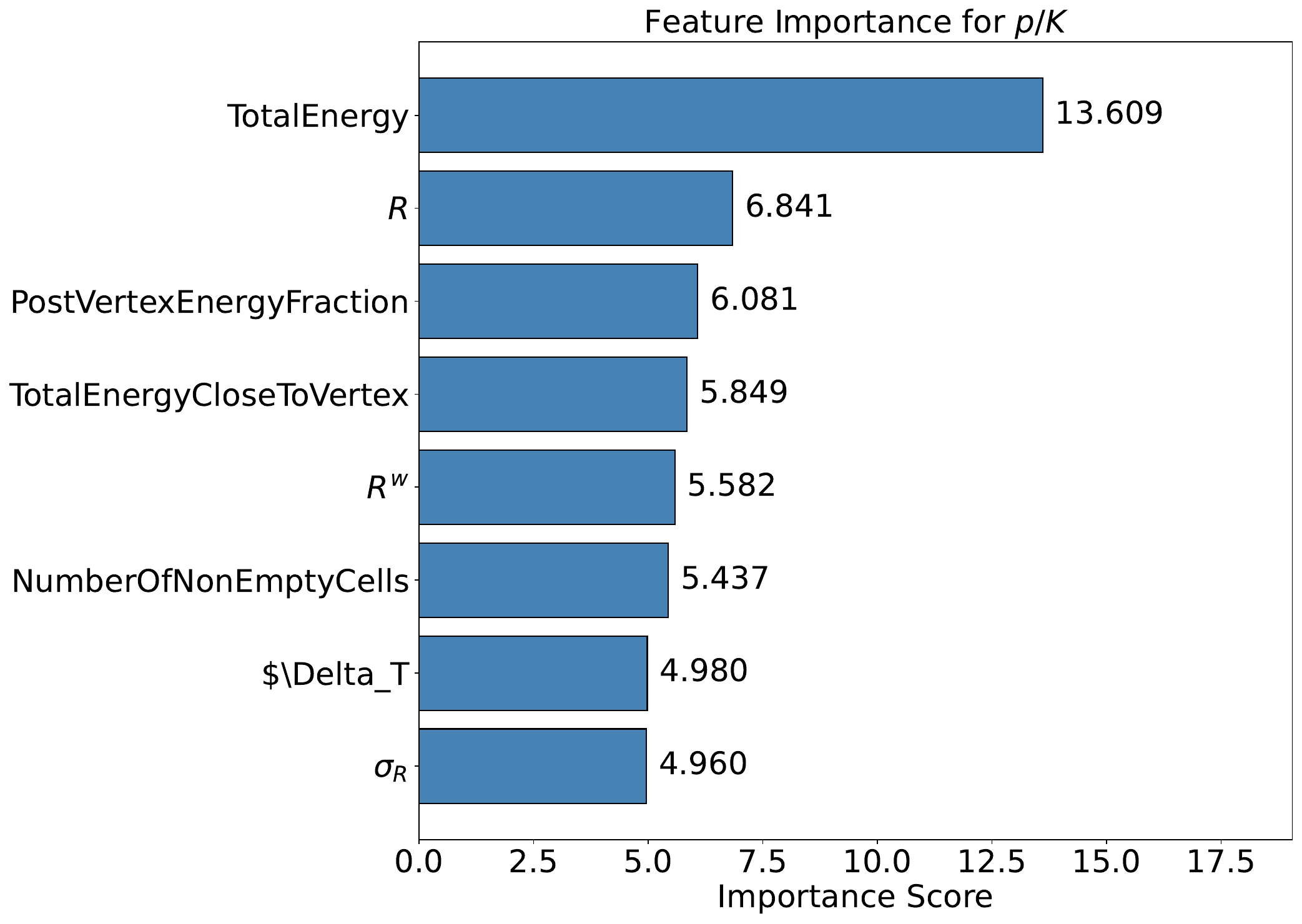} 
        \caption*{}
        \label{fig:feature_importance_pk}
    \end{subfigure}
    \caption{Analysis with the XGBoost model for the $p/K$ classification (cell size of $3 \times 3 \times 12 \, \text{mm}^3$). (Left) Distribution of the winning probability, categorized into correct classification (blue) and incorrect classification (red). (Right) Top 8 most important features used by the XGBoost model in the $p/K$ classification, ranked in descending order of significance. }
    \label{fig:combined_results_pk}
\end{figure}

This third analysis provides results that can be considered intermediate to the previous two analyses. By examining the distribution of winning probabilities (see Fig.~\ref{fig:combined_results_pk}, left), it can be seen that there is a threshold value above which the distribution of correct predictions surpasses the one of incorrect predictions. However, this result is not as advantageous as that described in Sec.~\ref{sec:xgboost_ppi}, since it would result in a significant loss of efficiency.

\begin{figure}[H]
    \centering
    \includegraphics[width=0.8\textwidth]{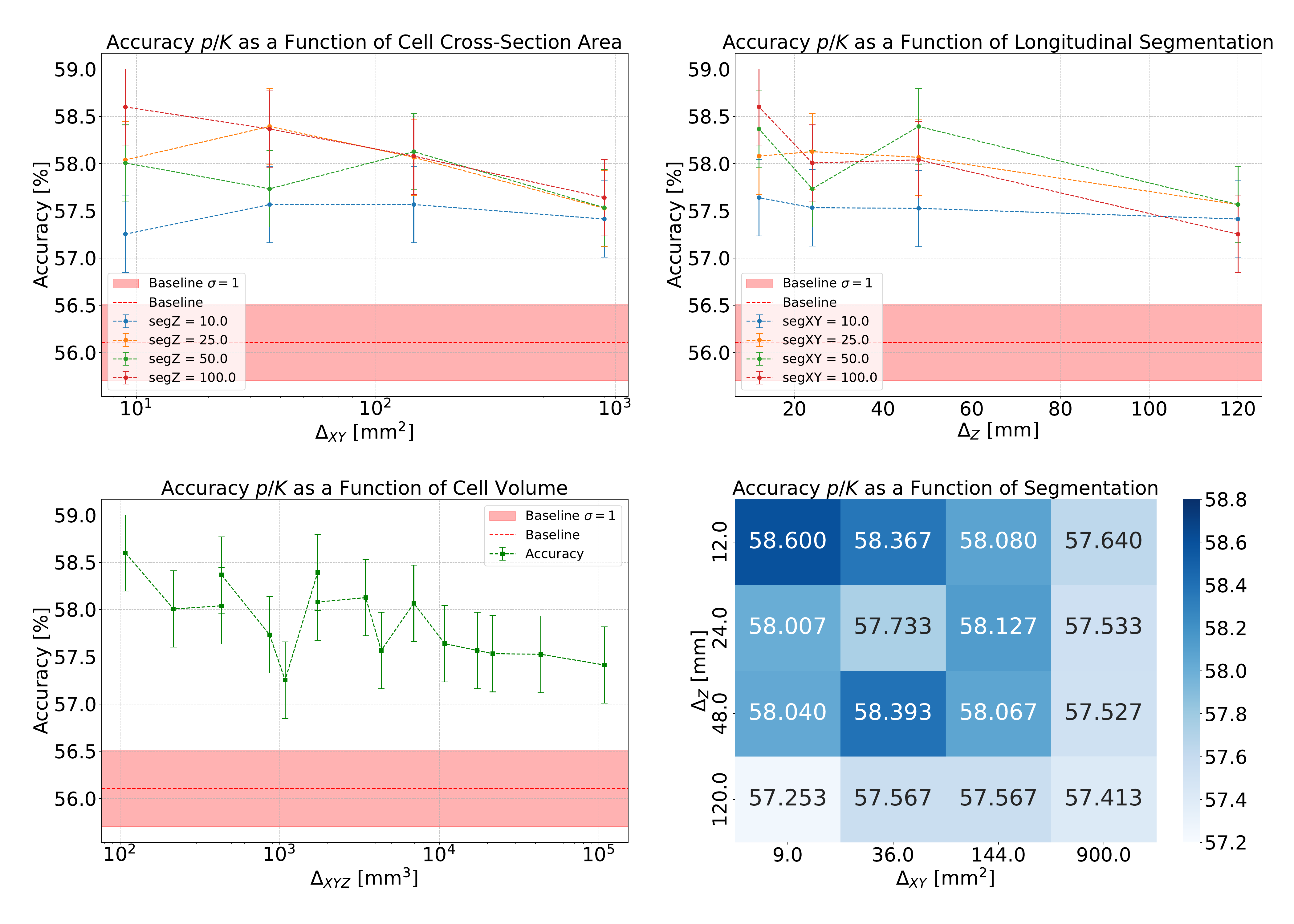} %
    \caption{Summary results on $p/K$ classification with XGBoost. Studies on the dependence of the accuracy on cell cross section, longitudinal segmentation and volume.  Because of the use of the same input data, the reported accuracy values are correlated. }
    \label{fig:moneyplot_p_k_xgboost}
\end{figure}

In Fig.~\ref{fig:combined_results_pk} right, it can be seen that the total energy released dominates among the input features, suggesting that the impact of segmentation in this case is less significant. Finally, we observe the trend of accuracy with respect to the baseline and its dependence on cell size. In the bottom left part of Fig.~\ref{fig:moneyplot_p_k_xgboost}, a decreasing trend is visible with increasing cell volume, and the same trend can be observed for both transverse and longitudinal segmentation (see Fig.~\ref{fig:moneyplot_p_k_xgboost}, top).


\subsection{Deep Neural Network}

This section discuss the results obtained using Deep Neural Network (DNN). The DNN model is trained on the NVIDIA GeForce RTX 4090 GPU, completing 180 epochs in approximately 25 minutes.

\begin{figure}[H]
    \centering
    \includegraphics[width=0.7\textwidth]{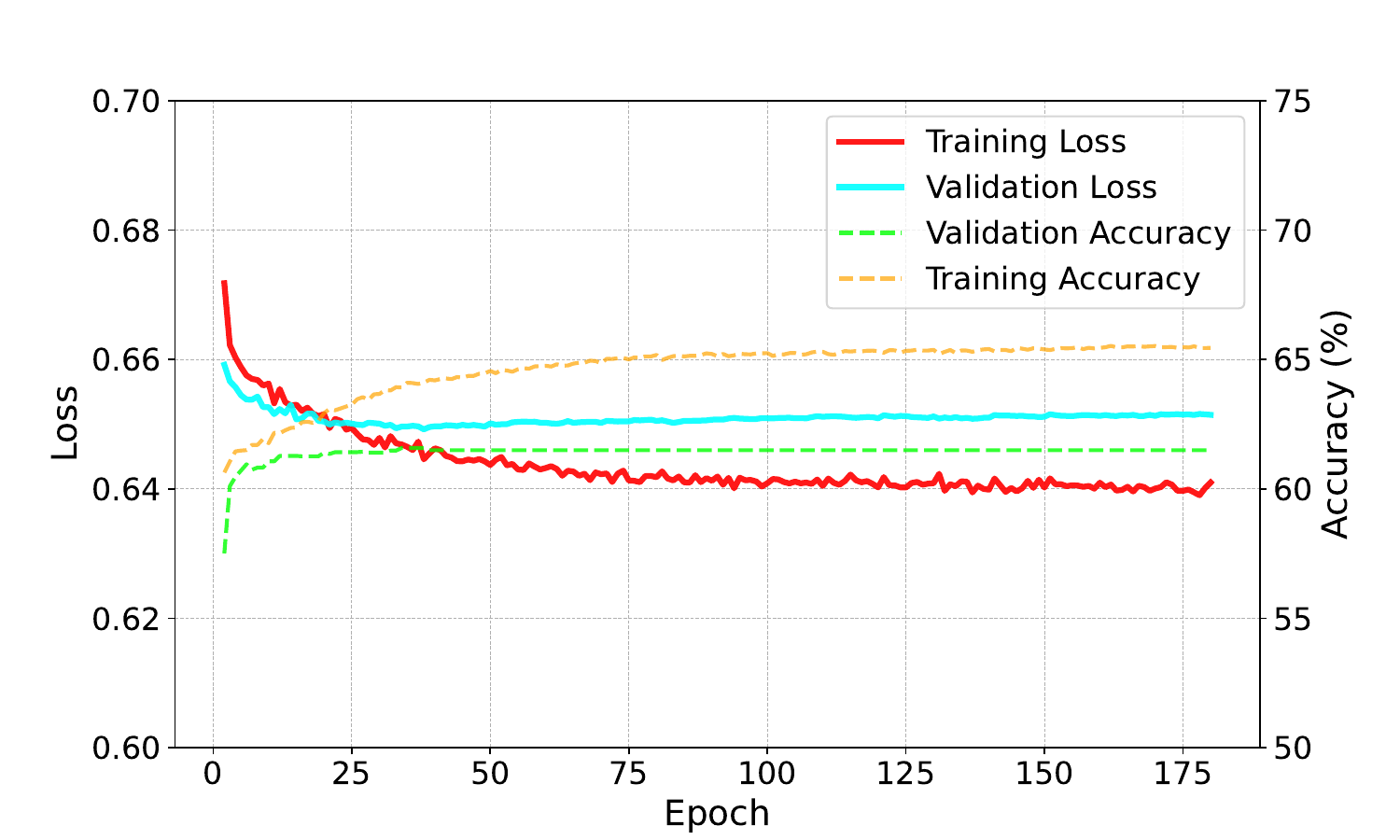} %
    \caption{Training and validation accuracy, along with training and validation loss.}
    \label{fig:lr_acc}
\end{figure}

Fig.~\ref{fig:lr_acc} shows the training and validation loss, as well as the training and validation accuracy. It is evident that both the loss and accuracy saturate within the range of epochs studied.

\begin{figure}[H]
        \centering
        \includegraphics[width=0.8\textwidth]{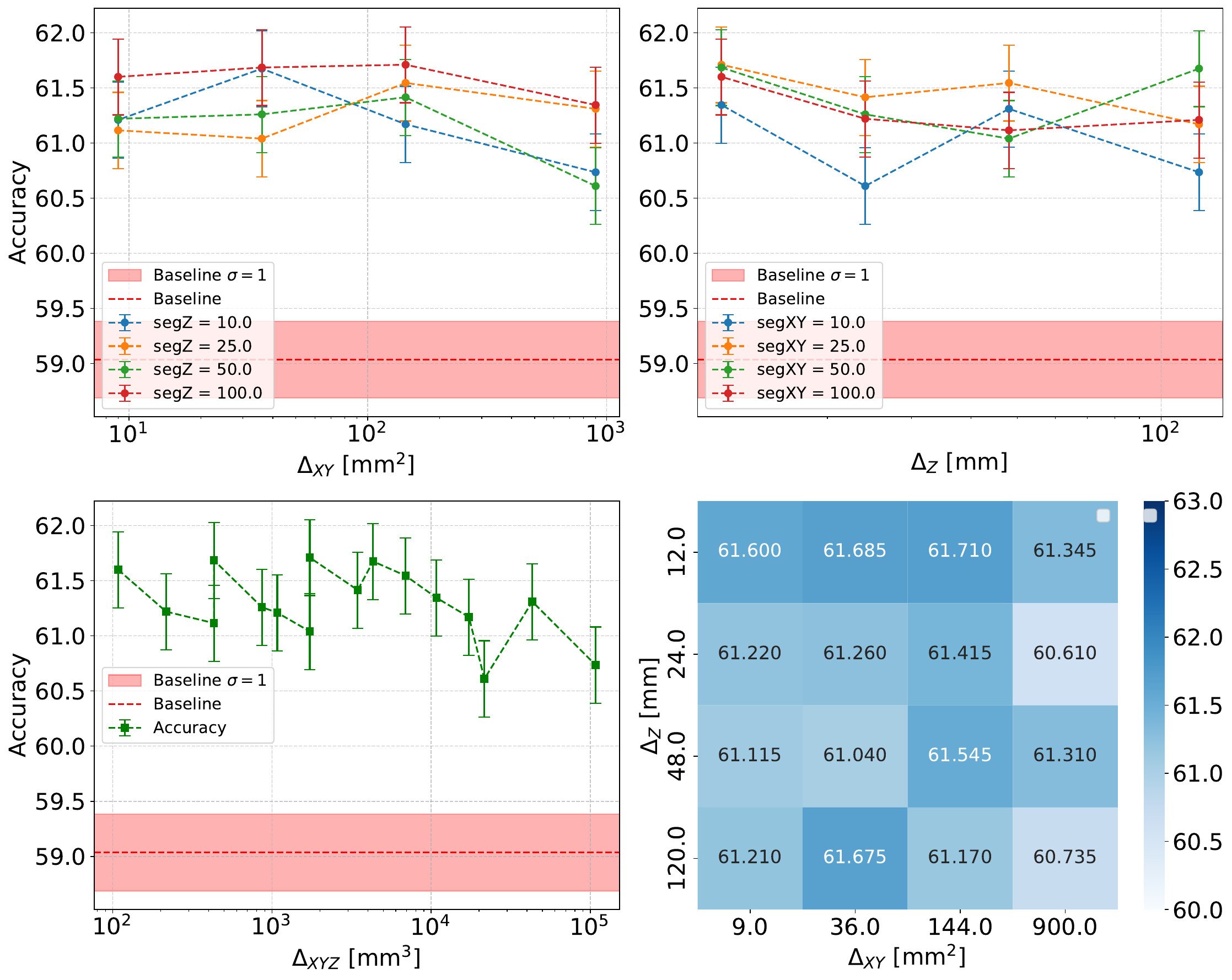}
        \caption{Summary of classification results for the $p/\pi$ with DNN. (Top-Left) Variation in accuracy with changes in transverse (XY) segmentation for a highly granular calorimeter while fixing longitudinal segmentation (Z). (Top-Right) Accuracy as function of the longitudinal segmentation while keeping the XY segmentation constant. (Bottom-Left) Accuracy variations with changes in the overall volume of calorimeter cells. (Bottom-Right) Accuracy matrix, summarizing accuracy values for different datasets. Because of the use of the same input data, the reported accuracy values are correlated. }
        \label{fig:moneyplot_p_pi_dnn}
\end{figure}
Fig.~\ref{fig:moneyplot_p_pi_dnn} show the results obtained using a Deep Neural Network (DNN) for the classification of $p/\pi$. The plots illustrate how accuracy varies with changes in the granularity of the calorimeter. Uncertainty bars in the DNN graphs of accuracy are computed using Clopper-Pearson interval with $\alpha=0.32$ using Eq.~\ref{eqn:CI}. See Sec.~\ref{sec:metrics} for more details.
Along with $p/\pi$ classification,DNN model for the classification between $\pi/K$  and $p/K$ has been tested. 

\begin{figure}[t]
    \centering
    \begin{subfigure}[b]{0.45\textwidth}
        \centering
        \
        \includegraphics[width=\textwidth]{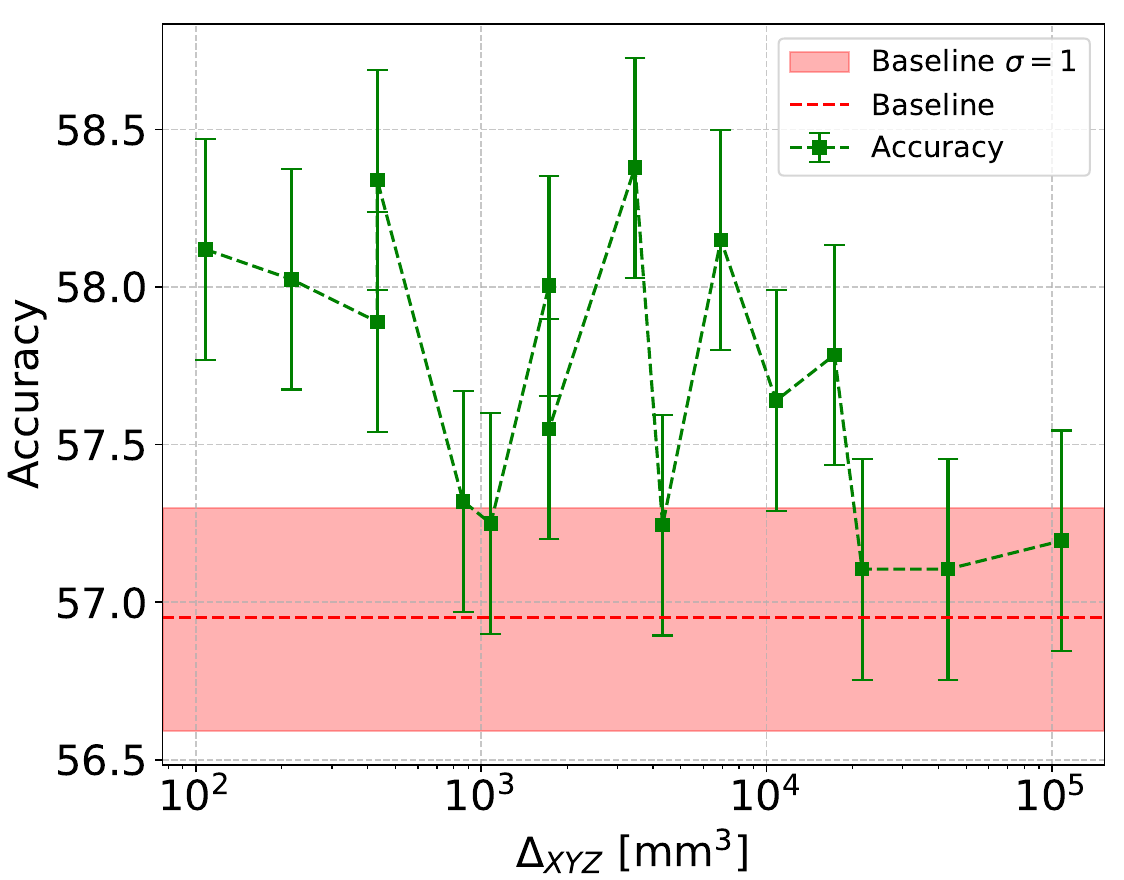}
        \captionsetup{justification=centering}
        \caption{}
       
    \end{subfigure}
    \hfill
    \begin{subfigure}[b]{0.45\textwidth}
        \centering
        \includegraphics[width=\textwidth]{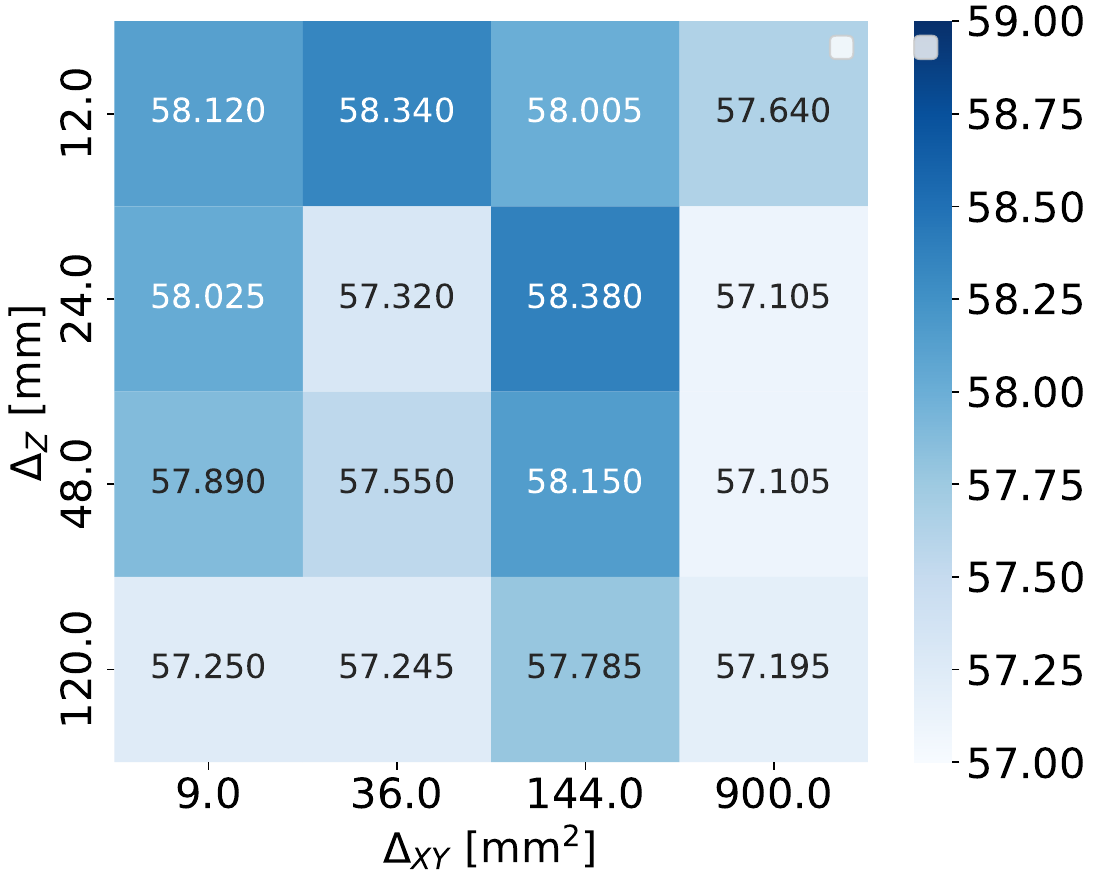}
        \captionsetup{justification=centering}
        \caption{}
      
    \end{subfigure}
    \caption{$p/K$ classification results using DNN:~(a) Accuracy variation with changes in cell volume~(b) Accuracy matrix for different cell dimensions.  Because of the use of the same input data, the reported accuracy values are correlated. }
    \label{fig:pk_money}
\end{figure}

\begin{figure}[H]
    \centering
    \begin{subfigure}[b]{0.48\textwidth}
        \centering
        \
        \includegraphics[width=\textwidth]{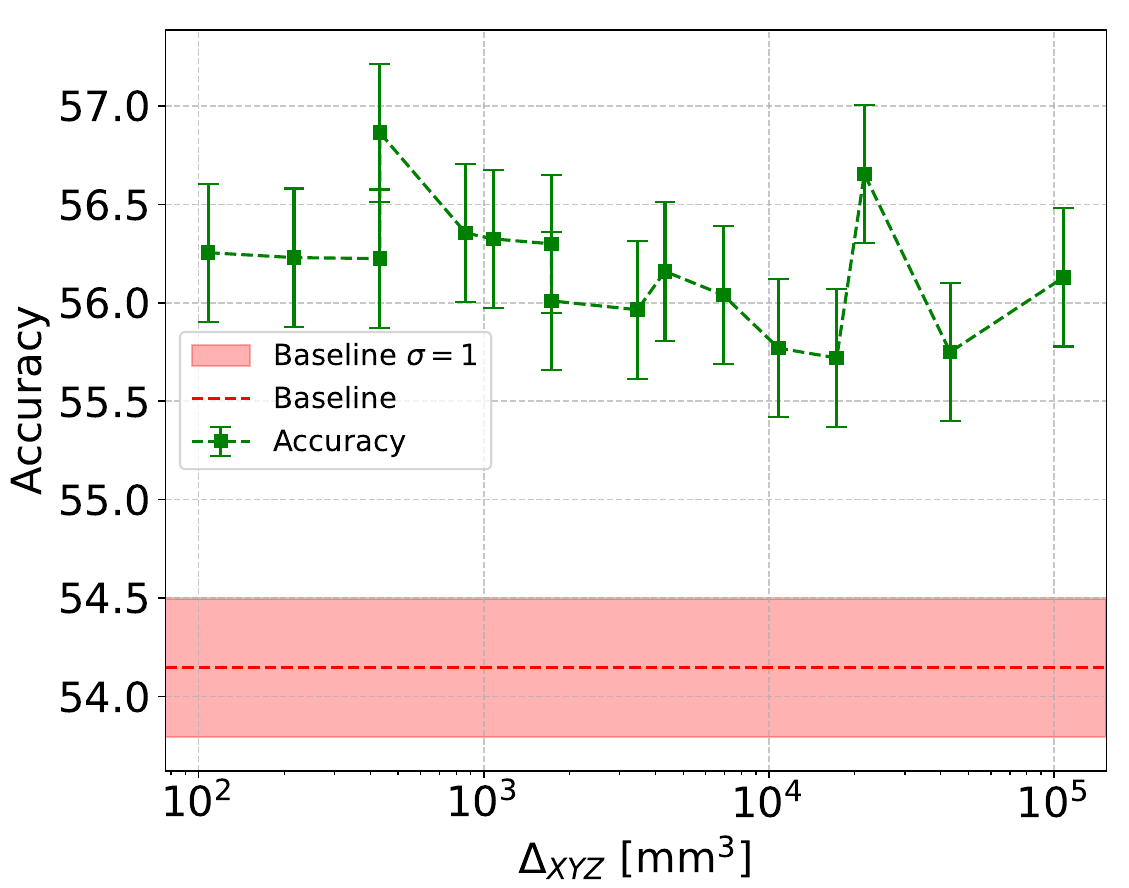}
        \captionsetup{justification=centering}
        \caption{}
       
    \end{subfigure}
    \hfill
    \begin{subfigure}[b]{0.48\textwidth}
        \centering
        \includegraphics[width=\textwidth]{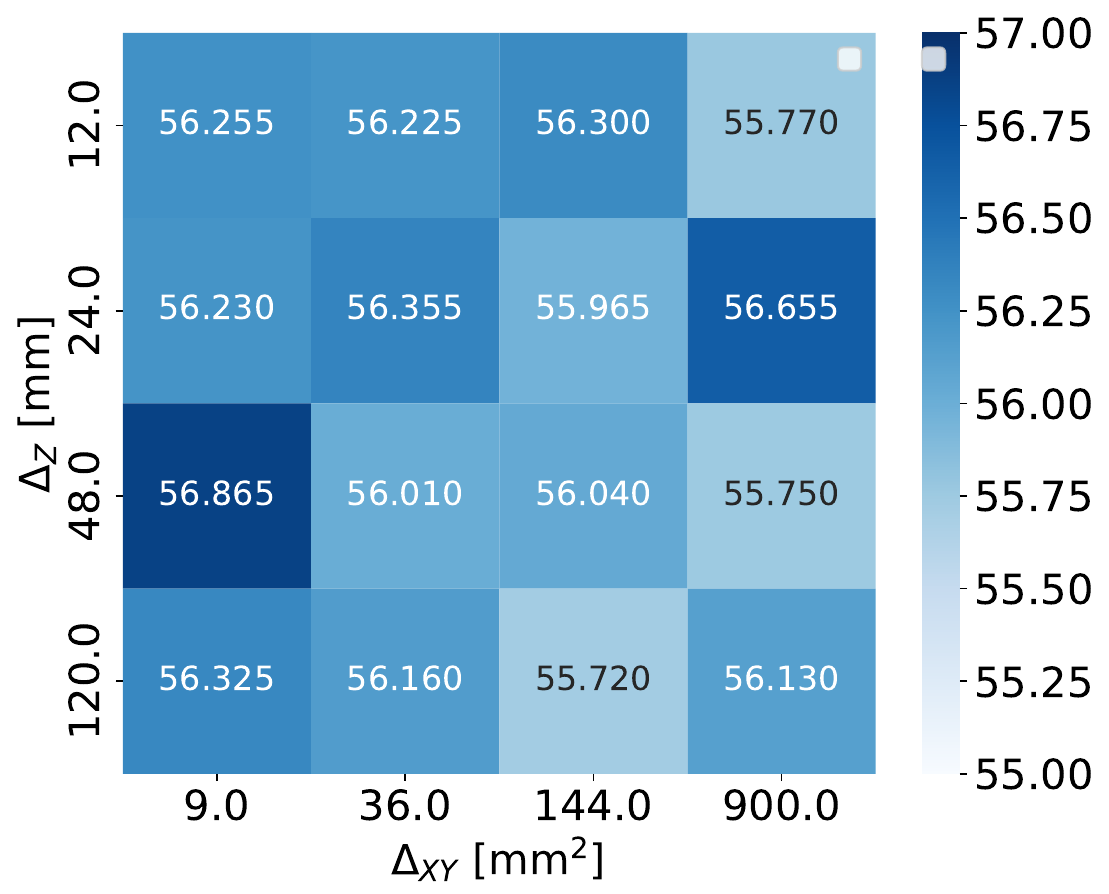}
        \captionsetup{justification=centering}
        \caption{}
      
    \end{subfigure}
    \caption{$\pi/K$ classification results using DNN:~(a) Accuracy variation with changes in cell volume~(b) Accuracy matrix for different cell dimensions.  Because of the use of the same input data, the reported accuracy values are correlated. }
    \label{fig:pik_money}
\end{figure}

Figures~\ref{fig:pk_money} and~\ref{fig:pik_money} illustrate the relationship between granularity and accuracy for $p/K$ and $\pi/K$ classifications, respectively. A slight decrease in accuracy is observed as the cell volume increases. However, this decline is not monotonic, and considering the error bars, the variation appears minimal. The specific reasons for this behavior are discussed in Sec.~\ref{sec:xgboost_ppi},~\ref{sec:xgboost_pk} and~\ref{sec:xgboost_pik}.

When comparing the performance of the XGBoost model and Deep Neural Networks (DNN), it is observed that both models yield similar results, effectively eliminating the possibility of selection bias. However, XGBoost stands out as a more feasible and cost-effective option in terms of computational efficiency. While DNNs are highly effective for tensor-based data, the tabular data structure used in this context aligns better with the strengths of XGBoost. 
Additionally, XGBoost offers a simpler implementation and has the advantage of being an interpretable model, as it provides explicit feature importance. In contrast, DNNs determine feature importance implicitly, making their interpretability more challenging. This makes XGBoost not only a computationally efficient choice but also a model that offers greater insights into the data.

\section{Related Work}
\label{sec:RelatedWorks}

High-Granularity Calorimeters will be the natural choice in future particle physics experiments at high-luminosity colliders, where the number of produced particles in each interaction may further increase from the already large value they have in today's LHC collisions. This will generate a large amount of data, which is difficult to process using traditional methods due to the high computational demands. Recently, machine learning algorithms have become useful tools for handling this data, especially for tasks like particle classification and regression. For example, as shown in~\cite{belayneh_calorimetry_2020}, neural networks have demonstrated significant progress in energy regression and particle classification for HGCAL. They simulated CMS and ATLAS calorimeter geometry and used both electromagnetic and hadron calorimter to seperate $\gamma$/$\pi^0$ and $e^-$/$\pi^\pm$. The GlueX experiment at Jefferson Lab used neural networks to separate background photons from hadron interactions and signal photons from $\omega$-meson decays. This highlights the power of neural networks in rejecting background noise, especially for neutral particles~\cite{Barsotti_2020}. Many existing methods employ Convolutional Neural Networks (CNNs), which are well-suited for image-based data, including visualizations of calorimetric showers. However, calorimetric showers inherently lack a natural ordering, unlike images which are structured grids. This unordered nature of calorimetric data makes point cloud representations a more suitable and intuitive choice.

As demonstrated in~\cite{Torales_Acosta_2024}, point cloud representations effectively capture the spatial and energy distribution of calorimetric showers. These representations enable the use of permutation-invariant architectures like DeepSets~\cite{zaheer2018deepsets}, which are specifically designed to handle unordered data. This approach allows for a more natural modeling of calorimetric showers and has been successfully applied to accelerate their simulation, offering both computational efficiency and accuracy compared to traditional methods.

\section{Conclusions}
\label{sec:Conclusions}

High granularity is today an important requirement for calorimeters in high-energy physics applications, due to the possibility to identify sub-jets from the decay of hadronically decaying, boosted heavy particles within wide jets, as well as to due to the benefits of the use of particle-flow techniques in event reconstruction.
In this work we have set out to study the possibility of distinguishing charged hadrons by the topological and time structure of their energy deposition in a homogeneous calorimeter of extremely high granularity, as a complementary piece of information that together with the requirements of boosted-jet tagging and particle-flow reconstruction may better inform the optimal design of future instruments. By studying a lead tungstate calorimeter block impinged by protons, pions, and kaons of positive charge and of 100 GeV of energy, we studied the distinguishability of these three classes through high-level features constructed with the information available from a hypothetical readout of energy and time of energy deposition in individual cells of very small size. By progressively merging cells into larger units we tried to ascertain how the harvested information would degrade in a less granular calorimeter. 

Our results reflect the current status of our investigations; they are still preliminary, as they do not conclusively ascertain yet what is the attainable, ultimate state-of-the-art level of discrimination of the different particle species, nor the scale of cell volume below which the required information is preserved. To achieve those goals it will be necessary to couple the high-level feature approach with a convolutional neural network or graph-based network, which can extract more information from raw detector outputs; the combined information of low-level and high-level information is guaranteed to offer results of closer to ultimate power. In addition, a study of a full spectrum of different particle energies in the range typical of collider physics applications [1 GeV - 1TeV], and consideration of negatively-charged particles (antiprotons, negative pions and kaons), neutral ones (neutrons, $K^0_L$) as well as deuterons and anti-deuterons will provide a better overall picture of the use of granular calorimeters for event reconstruction and particle identification. We intend to pursue the above studies in future work; regardless of the {\it interim} nature of presented results, we believe they  already show how useful particle identity information is extractable from the construction of informative high-level features that summarize the properties of the patterns of deposited energy and their time structure.


\dataavailability{The resources used for the analysis can be found on the following GitHub page: \href{https://github.com/andread3vita/TowardPIDwithGranularCalorimeters}{andread3vita/TowardPIDwithGranularCalorimeters}.}

\funding{
The work by TD and FS was partially supported by the Wallenberg AI, Autonomous Systems and Software Program (WASP) funded by the Knut and Alice Wallenberg Foundation.
The work by MA and FS was partially supported by the Jubilee Fund at the
 Lule{\aa} a University of Technology.
The work by PV was supported by the ``Ramón y Cajal” program under Project No. RYC2021-033305-I funded by the MCIN MCIN/AEI/10.13039/501100011033 and by the European Union NextGenerationEU/PRTR.
JK is supported by the Alexander-von-Humboldt foundation.

}

\appendixtitles{no}  
\appendixstart
\appendix
\section[]{First Nuclear Interaction Vertex Finder}
\label{sec:appendixA}

\noindent The \texttt{First Nuclear Interaction Vertex Finder} algorithm is designed to identify the index of the first "peak" in a vector of values, based on a specified threshold criterion. Below is a detailed breakdown of its components and functionality.

\subsection{Inputs And Parameters}
\noindent The function takes the following inputs:
\begin{itemize}
    \item \texttt{energyCoordinates}: A vector representing the spatial energy coordinates.
    \item \texttt{energyDeposition}: A vector representing the energy deposited at each coordinate.
    \item \texttt{threshold}: The initial threshold used to detect the peak in the energy profile.
\end{itemize}

\subsection{Step-By-Step Algorithm Overview}
\subsubsection*{\textbf{Energy Profile Calculation}}
\noindent The algorithm processes the interactions to construct an energy profile along the z-dimension:

\begin{enumerate}
    \item Spatial coordinates and energy deposition (\texttt{E}) are retrieved for each interaction.
    \item Energy contributions are accumulated into z-slices within the XY window range.
    \item To improve peak detection, the energy profile is smoothed using a moving average filter.
\end{enumerate}

\subsubsection*{\textbf{First Pass: Initial Peak Detection}}
\noindent In the first pass, the algorithm scans the \texttt{energyProfile} to locate the first peak using the original threshold:
\begin{enumerate}
    \item Iterate through the elements of the energy profile:
    \begin{lstlisting}
    for (int i = 0; i < size; ++i) {
        if (energyProfile[i] > threshold) {
            return i;
        } else if (energyProfile[i + 1] - energyProfile[i] > threshold) {
            return i + 1;
        }
    }
    \end{lstlisting}
    
    \item The algorithm identifies significant peaks in a sequence by comparing the values of consecutive elements. If an element surpasses a predefined threshold, its index is immediately returned as a peak. Alternatively, if the difference between two consecutive elements exceeds the threshold, the index of the second element is returned, marking it as the peak.
\end{enumerate}

\subsubsection*{\textbf{Second Pass: Threshold Reduction}}
\noindent If no peak is found in the first pass, the threshold is gradually reduced, and the search is repeated:

\begin{enumerate}
    \item Decrease the threshold by 10\% in each iteration:
    \begin{lstlisting}
    for (int j = 1; j < 5; ++j) {
        double thr = threshold - (j * 0.1) * threshold;
    }   
    \end{lstlisting}
    
    \item Perform the peak detection search with the new threshold:
    \begin{lstlisting}
    for (int i = 0; i < size; ++i) {
        if (energyProfile[i] > thr) {
            return i;
        } else if (energyProfile[i + 1] - energyProfile[i] > thr) {
            return i + 1;
        }
    }
    \end{lstlisting}
\end{enumerate}

\subsubsection*{\textbf{Handling Cases With No Peak}}
\noindent If no peak is detected after both passes, the function returns \texttt{-1}, indicating that no significant peaks were found in the vector.

\subsection{Summary}
\begin{itemize}
    \item The function employs a greedy approach, returning the index of the first detected peak in the energy profile.
    \item By gradually reducing the threshold, the function becomes more sensitive to smaller variations in the data, improving peak detection in cases of low-energy deposition.
    \item If no peak is identified after both search passes, the function returns \texttt{-1}, indicating that no peak meets the specified criteria.
\end{itemize}

\subsection{Performance Evaluation}

\noindent To evaluate the algorithm's performance, an accuracy metric is defined. This metric corresponds to the ratio between the number of events where the absolute difference between the trueZvertex and the recoZvertex is less than 2, and the total number of events. Based on this definition, the algorithm's performance was analyzed for various longitudinal segmentations.
\begin{figure}[H]
    \centering
    \includegraphics[width=0.55\textwidth]{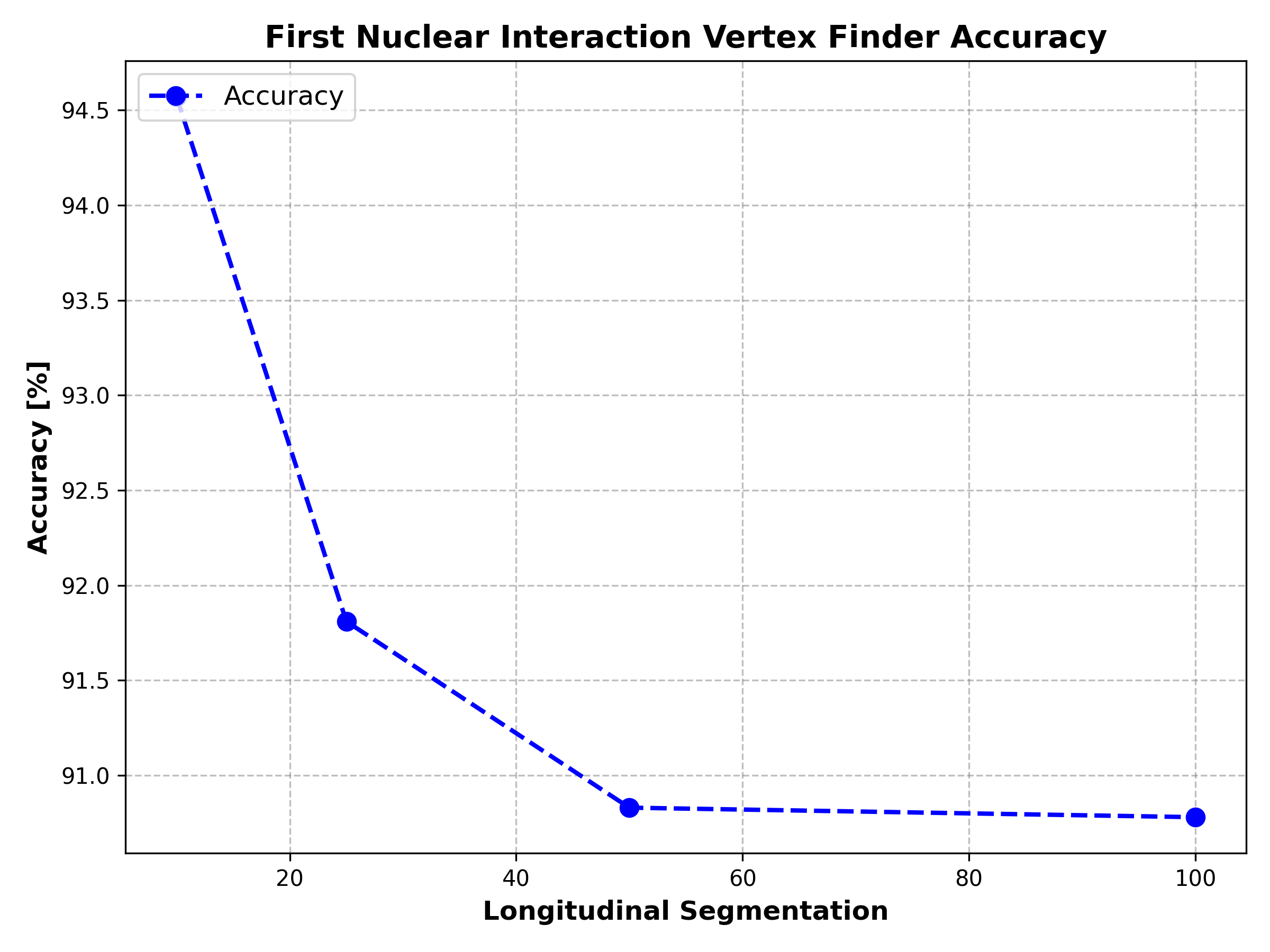} %
    \caption{Accuracy of First Nuclear Interaction Vertex Finder as a function of the longitudinal segmentation.}
    \label{fig:algorithm_performance}
\end{figure}
 As shown in Fig.~\ref{fig:algorithm_performance}, the accuracy is always above 90\%, with a minimum value of 90.78\% when the segmentation is set to 100.
\section[]{Energy Peak Finder}
\label{sec:appendixB}
\noindent The \texttt{Energy Peak Finder} function is designed to analyze events, identifying the most significant peak in the energy deposition profiles along the X, Y, and Z axes. Below is a step-by-step explanation of the function and its operations.

\subsection{Inputs And Parameters}
\noindent The function accepts the following parameters:
\begin{itemize}
    \item \texttt{sphere\_radius}: A parameter that defines the radius of the sphere centered on the cell containing the energy peak.
\end{itemize}

\subsection{Step-by-Step Algorithm Overview}

\subsubsection*{\textbf{Histograms Creation}}

\noindent Two 2D histograms (\texttt{hist\_cell\_zy} and \texttt{hist\_cell\_zx}) are created to represent energy deposits in the Z-Y and Z-X planes, respectively. These histograms are filled on the basis of the event data.

\subsubsection*{\textbf{Energy Profile Along Z-Axis And Peak Detection}}
\noindent The energy deposition data are projected along the Z-axis:
\begin{enumerate}
    \item A projection of the \texttt{hist\_cell\_zy} histogram onto the Z axis is stored in \texttt{projZ}.
    \item \texttt{TSpectrum::Search} is used to find peaks in \texttt{projZ} with a threshold of 0.1.\cite{smooth}
    \item The positions of the detected peaks along the Z-axis are stored in \texttt{peaksZ}.
    \item The peaks are sorted in increasing order of associated energy, and the highest  energy one is finally stored.
\end{enumerate}

In most analyzed events the algorithm finds a single peak; cases when multiple peaks compete for being classified as the first event vertex are very rare; the highest-energy one is anyway used.

\subsubsection*{\textbf{Peak Filtering And Search In X And Y Projections}}
\noindent Given the Z peak position, the function performs the following steps:
\begin{enumerate}
    \item Filters the \texttt{hist\_cell\_zx} and \texttt{hist\_cell\_zy} histograms based on the Z peak position.
    \item Projects the filtered histograms onto the X and Y axes, respectively, creating \texttt{projX} and \texttt{projY}.
    \item Searches for peaks in the X and Y projections using \texttt{TSpectrum::Search}.\cite{smooth}
    \item If multiple peaks are found in X or Y, the algorithm selects the most prominent peak by comparing the peak intensities.
\end{enumerate}

\subsubsection*{\textbf{Energy Accumulation Around Peaks}}
\noindent The function accumulates the energy deposition values around the detected peak:
\begin{itemize}
    \item For each energy deposition in the event, the 3D position is converted to cell coordinates.
    \item The proximity of the energy deposition to the detected peaks is evaluated, and the energy is added if the energy deposition is within the sphere defined by \texttt{sphere\_radius}.
\end{itemize}

\subsection{Results}
\noindent Finally, the function returns a vector that contains the energy and the energy ratio of the most significant peak.

\section[]{Feature Distributions For Proton, Pion And Kaons}
\label{sec:appendixC}
\begin{figure}[H]
\hspace{-4cm}
    \begin{minipage}[t]{1.3\textwidth}
        \noindent
        \includegraphics[width=1.\textwidth]{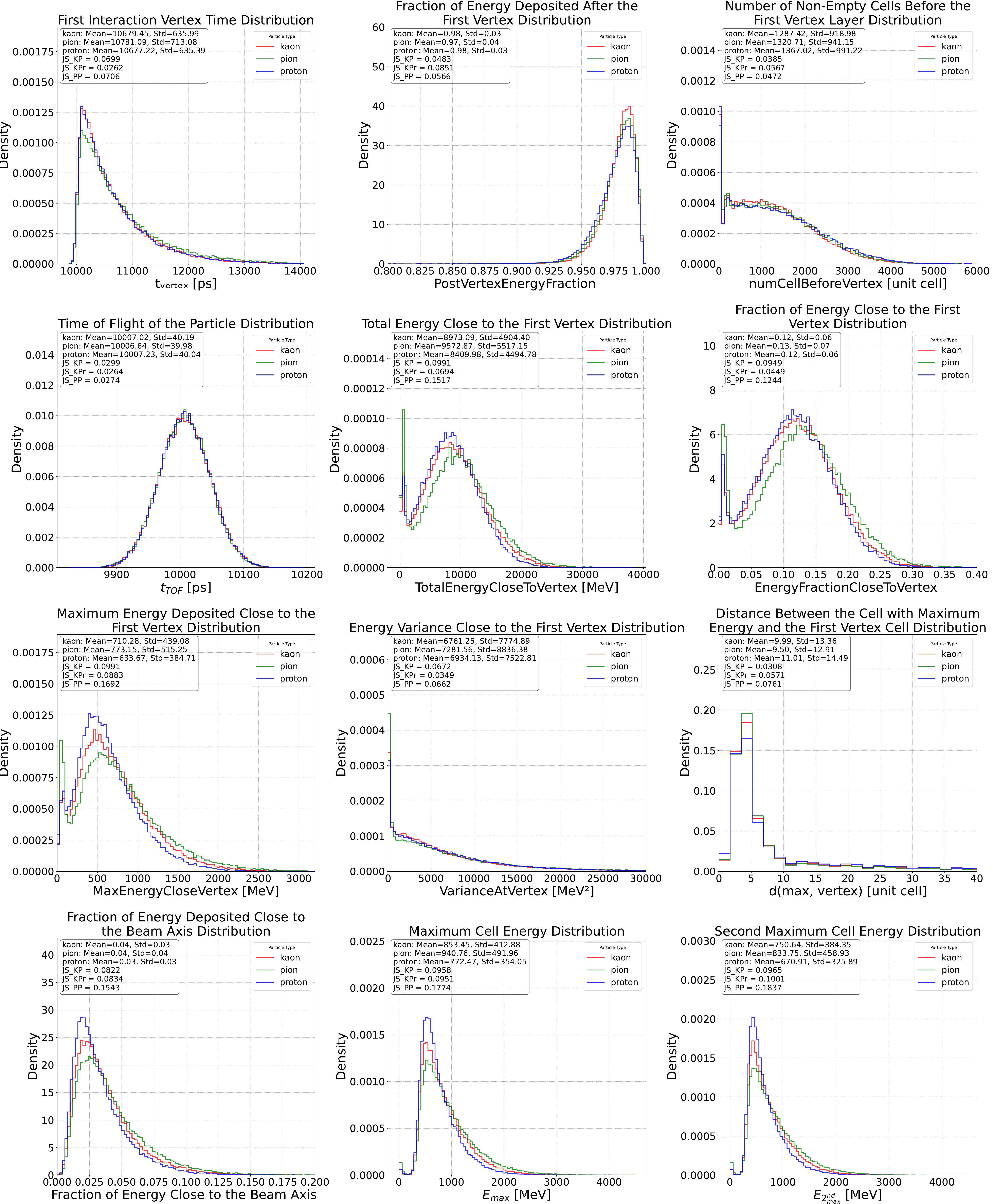}
     
        \caption{These figures present the distributions of all features used in the analysis (cell size of $3 \times 3 \times 12 \, \text{mm}^3$). The results are shown for each particle, and the Jensen-Shannon Divergence is reported for each particle pair to quantify the similarity between their respective distributions.}
    \end{minipage}
\end{figure}

\begin{figure}[H]
\hspace{-4cm}
    \begin{minipage}[t]{1.3\textwidth}
        \noindent
        \includegraphics[width=1.\textwidth]{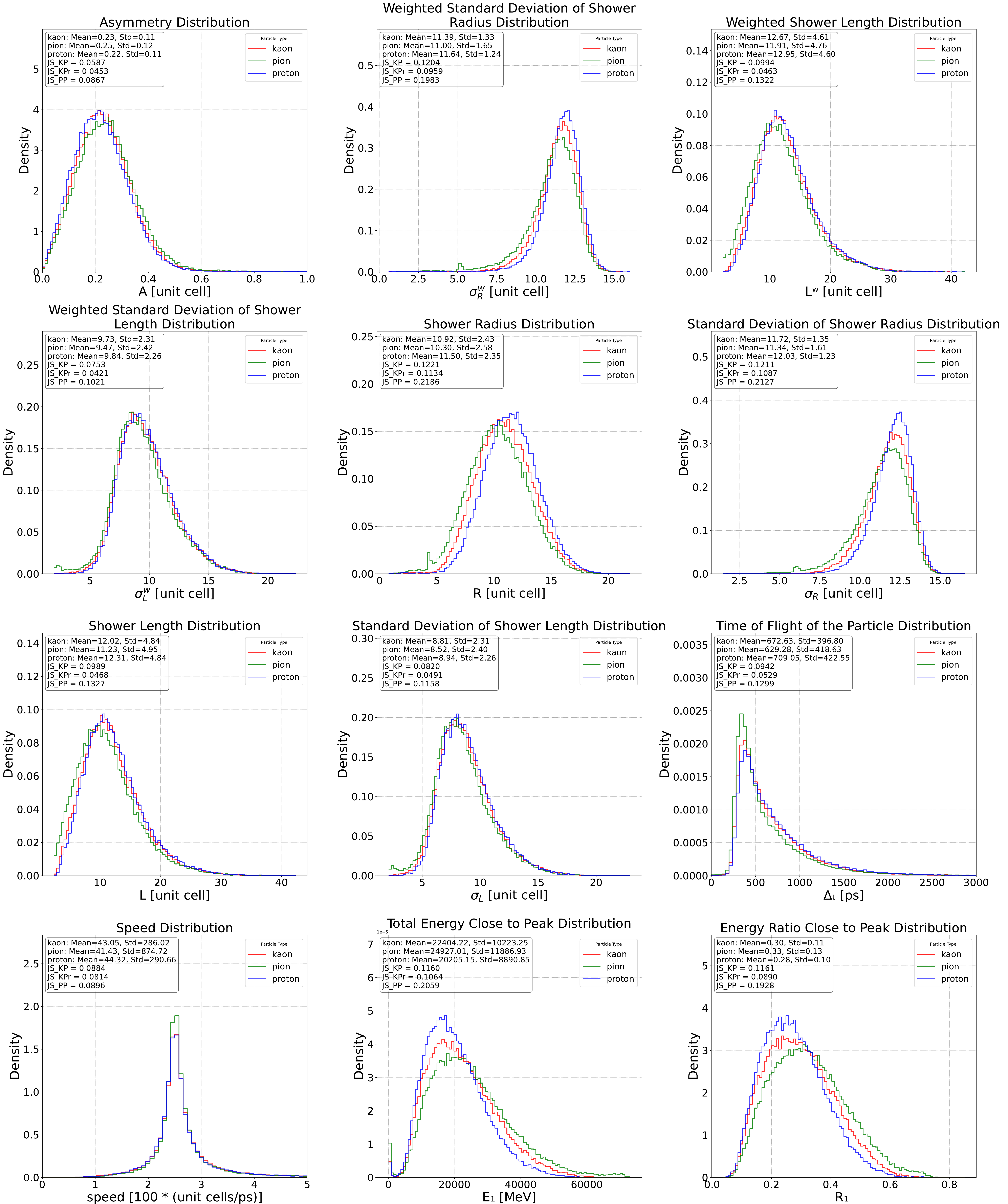}
     
        \caption{These figures present the distributions of all features used in the analysis (cell size of $3 \times 3 \times 12 \, \text{mm}^3$). The results are shown for each particle, and the Jensen-Shannon Divergence is reported for each particle pair to quantify the similarity between their respective distributions.}
    \end{minipage}
\end{figure}

\begin{figure}[H]
\hspace{-4cm}
    \begin{minipage}[t]{1.3\textwidth}
        \noindent
        \includegraphics[width=1.\textwidth]{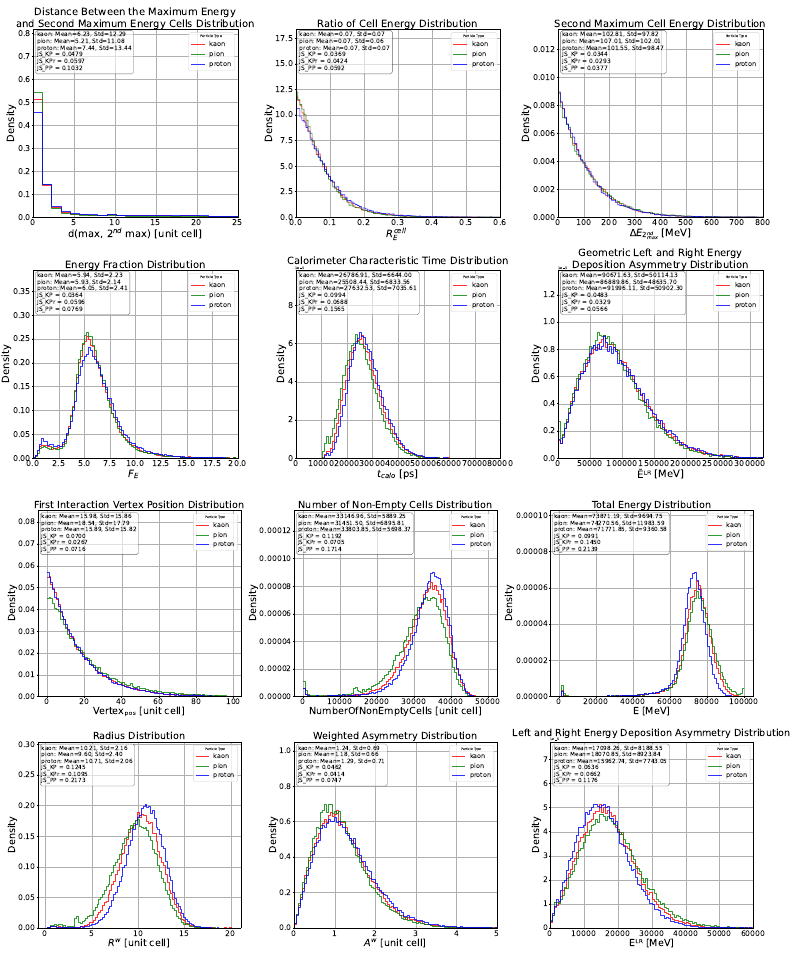}
     
        \caption{These figures present the distributions of all features used in the analysis (cell size of $3 \times 3 \times 12 \, \text{mm}^3$). The results are shown for each particle, and the Jensen-Shannon Divergence is reported for each particle pair to quantify the similarity between their respective distributions.}
    \end{minipage}
\end{figure}

\newpage


\begin{adjustwidth}{-\extralength}{0cm}

\reftitle{References}
\bibliography{main.bib}



\PublishersNote{}
\end{adjustwidth}
\end{document}